\documentclass[aps,prd,twocolumn,nofootinbib]{revtex4-1} %nofootinbib
\usepackage{color}
\usepackage[latin1]{inputenc}
\usepackage{float,amsmath}
\usepackage{graphicx}
\usepackage[normalem]{ulem}

\newcommand{\xt}{{\mathbf{x}_\perp}}
\newcommand{\kt}{{\mathbf{k}_\perp}}

\begin{document}

\title{Azimuthal anisotropies in p+Pb collisions from classical Yang-Mills dynamics}

\author{Bj\"orn Schenke}
\affiliation{Physics Department, Brookhaven National Laboratory, Upton, NY 11973, USA}
\author{S\"oren Schlichting}
\affiliation{Physics Department, Brookhaven National Laboratory, Upton, NY 11973, USA}
\author{Raju Venugopalan}
\affiliation{Physics Department, Brookhaven National Laboratory, Upton, NY 11973, USA}

\begin{abstract}
We compute single and double inclusive gluon distributions in classical Yang-Mills simulations of proton-lead collisions and extract the associated transverse momentum dependent Fourier harmonics $v_2(p_T)$ and $v_3(p_T)$. Gluons have a large $v_2$ in the initial state, while odd harmonics such as $v_3$ vanish identically at the initial time $\tau=0^{+}$. By the time $\tau \lesssim 0.4\,{\rm fm/c}$ final state effects in the classical Yang-Mills evolution generate a non-zero $v_{3}$  and only mildly modify the gluon $v_{2}$. 
Unlike hydrodynamic flow, these momentum space anisotropies are uncorrelated with the global spatial anisotropy of the collision.
A principal ingredient for the generation of  $v_2$ and $v_3$ in this framework is the event-by-event breaking of rotational invariance in domains the size of the inverse of the saturation scale $Q_s$.
In contrast to our findings in p+Pb collisions Yang-Mills simulations of lead-lead collisions generate much smaller values of $v_{2,3} (p_T)$ and additional collective flow effects are needed to explain experimental data. This is because the locally generated anisotropy due to the breaking of rotational invariance is depleted with the increase in the number of uncorrelated domains.
\end{abstract}

%\pacs{11.15Bt, 04.25.Nx, 11.10Wx, 12.38Mh}
\maketitle

%%%%%%%%%%%%%%%%%%%%%%%%%%%%%%%%%%%%%%%%%%%%%%%%%%%%%%%%%%%%%%%

\section{Introduction}
A striking result from high-multiplicity proton-proton (p+p) and proton-lead (p+Pb) collisions at the LHC was the discovery of a ``double-ridge'' structure in two-particle correlations that is long range in their rapidity difference $\Delta \eta$ and includes a dominant $\cos(2\Delta\phi)$ modulation in their relative azimuthal angle~\cite{Khachatryan:2010gv,CMS:2012qk,Chatrchyan:2013nka,Aad:2012gla,Aad:2013fja,Aad:2014lta,Abelev:2012ola,ABELEV:2013wsa}. Key aspects of the structure of the observed correlations in proton-nucleus collisions bear a striking similarity to those observed in heavy-ion collisions \cite{Chatrchyan:2011eka,ALICE:2011ab,ATLAS:2012at} and may point to a form of collective behavior where many particles are correlated with each other.

In heavy-ion collisions the azimuthal structure of multi-particle correlations is quantitatively well described by viscous fluid dynamic calculations with fluctuating initial state geometries \cite{Gale:2012rq}. This naturally leads to the assumption that the physics responsible for the ridge structure in high multiplicity p+p and p+Pb events may also be driven by the final state collective flow of the system. Indeed, calculations using hydrodynamics (or microscopic models of final state interactions) are able to reproduce features of the azimuthal structure in p+Pb collisions \cite{Bozek:2013uha,Bozek:2013ska,Bzdak:2014dia,Qin:2013bha,Werner:2013ipa}.
Specifically, the Fourier coefficients $v_2(p_T)$ and $v_3(p_T)$ in the expansion of the particle distribution 
\begin{equation}
\label{eq:vnep}
  \frac{d N}{d \phi\,p_T\,dp_T} \propto 1 + 2 \sum_{n=1}^\infty v_n(p_T) \cos\left[ n \left(\phi-\psi_n(p_T)\right)\right]\, ,
\end{equation}
can be described by several of these models, albeit with different parameterizations of the initial state and transport coefficients. In Eq.~(\ref{eq:vnep}),  $\psi_n(p_T)= \frac{1}{n} \arctan \frac{\langle\sin(n\phi)\rangle_{\phi}}{\langle\cos(n\phi)\rangle_{\phi}}$ is the (transverse momentum dependent) event plane angle associated with the $n^{\rm th}$ harmonic, and $\langle\cdot\rangle_{\phi}$ denotes the azimuthal average with respect to the single inclusive particle distribution at a given $p_T$.

While some detailed features of the data --  such as the dependence of $v_2(p_T)$ on the particle mass -- can be explained quite naturally by final state collective effects, a number of conceptual problems can be identified within this theoretical approach. One concerns the applicability of viscous hydrodynamics due to large pressure gradients~\cite{Bzdak:2013zma,Niemi:2014wta} that are present for a significant fraction of the space-time evolution. Another concerns the sensitivity to the initial state in small systems~\cite{Schenke:2014zha}, which requires a better theoretical understanding of the early time dynamics. Further, the observation of pronounced azimuthal anisotropies even at high transverse momenta $p_T\gtrsim 3~{\rm GeV}$ challenges the hydrodynamic paradigm which is best applied to a description of low momentum excitations. It is therefore important to explore if multi-particle correlations at different transverse momentum scales can be understood in part or whole in alternative approaches. 

For instance, computations of intrinsic two particle correlations  in the Color Glass Condensate (CGC) framework have been shown to produce azimuthal anisotropies compatible with ridge data for even harmonics  for $p_T > 1$ GeV in p+p and p+Pb collision systems without the need for additional final state collective effects~\cite{Dumitru:2010iy,Dusling:2012iga,Dusling:2012cg,Dusling:2012wy,Dusling:2013oia}. However, no odd harmonics were generated because rescattering contributions to the intrinsic correlations were not included~\cite{Dusling:2014oha}. 
In addition to these intrinsic two-particle correlations, the presence of domains of directed chromo-electric fields inside the proton and the nucleus breaks rotational invariance on an event-by-event basis and thereby generates azimuthal anisotropies~\cite{Kovner:2012jm}. For quarks scattering off a colored target, these colored domains generate both $v_2$ and $v_3$~\cite{Dumitru:2014dra,Dumitru:2014vka,Noronha:2014vva}. However, gluons scattering off this target only generate even harmonics. We note that the possibility that azimuthal anisotropies could arise from the event-by-event breaking of rotational invariance by color fields has also been considered in a related approach \cite{Gyulassy:2014cfa}. 

Thus far the studies of initial state correlations generated by color-electric domains in \cite{Dumitru:2014dra,Dumitru:2014vka}, and recent extensions thereof in \cite{Lappi:2015vha}, have been based on a description of the proton as a dilute projectile of valence quarks scattering off the nucleus. However, experimentally significant anisotropies are only measured in events with \emph{very high multiplicities}, where it is more appropriate to describe both the proton and the lead nucleus as dense colored objects. Because of the high gluon occupancies in both the proton and the lead nucleus, they can be approximated as classical gluon fields, and their leading order space-time evolution is described by solving classical Yang-Mills equations.

Such classical Yang-Mills (CYM) simulations were performed previously for nucleus-nucleus collisions~\cite{Krasnitz:1999wc,*Krasnitz:2000gz,Lappi:2003bi} (including an early study of elliptic flow in \cite{Krasnitz:2002ng}) and proton-nucleus collisions~\cite{Krasnitz:2002mn,Schenke:2013dpa}. More recently, these studies were extended to include more realistic initial conditions in the IP-Glasma model~\cite{Schenke:2012wb,Schenke:2012fw}, which provides a satisfactory description of multiplicities in high energy hadron collisions~\cite{Schenke:2013dpa}. When combined with a hydrodynamic evolution (MUSIC), the IP-Glasma+MUSIC model has also been successfully applied to the description of azimuthal harmonics in nucleus-nucleus collisions~\cite{Gale:2012rq}; azimuthal anisotropies in p+Pb collisions were however largely underestimated~\cite{Schenke:2014zha}. 
 
While the azimuthal anisotropy in the aforementioned studies~\cite{Gale:2012rq,Schenke:2014zha} was generated via the hydrodynamic evolution of the system, the IP-Glasma model also includes fluctuations of color charges inside the projectile and target. Such fluctuations break rotational invariance on an event-by-event basis and lead to a momentum space anisotropy already present in the initial state. In this letter, we will study these azimuthal anisotropies of gluons in the initial state and during the early time dynamics of proton-nucleus and nucleus-nucleus collisions. While the effect in Pb+Pb collisions is small, we find that initial state and early time effects are sizable in p+Pb collisions and should be taken into account in the theoretical description of small collision systems.

This letter is organized as follows. In Sec.~II we briefly outline the theoretical framework underlying the classical Yang-Mills simulations. Our discussion follows the literature in the context of the IP-Glasma model~\cite{Schenke:2012fw} with the significant modification that we will also consider `eccentric' proton configurations following~\cite{Schlichting:2014ipa}. We then discuss the measurement of azimuthal anisotropies in this framework in Sec.~III and present numerical results for azimuthal Fourier harmonics of gluons in proton-nucleus collisions in Sec.~IV. We investigate the sensitivity of our results with respect to variations in the spatial color structure of the proton and perform a comparison of the effects in proton-nucleus and nucleus-nucleus collisions in Sec.~IV. The final section summarizes our conclusions and their implications for collective dynamics in proton-nucleus and nucleus-nucleus collisions.

\section{Theoretical framework}

Within the CGC framework \cite{Iancu:2002xk,Iancu:2003xm,Gelis:2010nm}, the dynamics of a high-energy collision is  -- to leading order in $\alpha_s$ -- described by solutions of the classical Yang-Mills equations,
\begin{equation}
\label{eq:YM1}
  [D_{\mu},F^{\mu\nu}] = J^\nu\,,
\end{equation}
in the presence of an eikonal color current $J^{\nu}$. Here, $D_{\mu}$ is the covariant derivative in the presence of the field $A_{\mu}$ and $F^{\mu\nu}$ is the gluon field strength tensor. For a right moving (projectile) proton and left moving (target) nucleus one has
\begin{equation}\label{eq:current}
  J^\nu = \delta^{\nu +}\rho_{\rm p}(x^-,\xt)+ \delta^{\nu -}\rho_{\rm Pb}(x^+,\xt)\,,
\end{equation}
and each event is characterized by a different color neutral distribution of random color charges $\rho_{\rm p/Pb}(x^{\mp},\mathbf{x})$ inside the proton and nucleus. 

Before the collision, the small-$x$ gluon fields inside the target and projectile nucleus are determined by the solution of (\ref{eq:YM1}) and can be compactly expressed\footnote{These expression are valid in light-cone gauge.} as~\cite{McLerran:1993ni,*McLerran:1993ka,*McLerran:1994vd,JalilianMarian:1996xn,Kovchegov:1996ty} 
\begin{eqnarray}
A^i_{\rm p/Pb}(\xt)=\frac{i}{g}V_{\rm p/Pb}(\xt)\partial^{i}V^{\dagger}_{\rm p/Pb}(\xt)\;,
\end{eqnarray}
in terms of the fundamental Wilson lines $V_{\rm p/Pb}(\xt)$ of the projectile and target nucleus. By dividing the longitudinal direction into $N_Y$ discrete rapidity intervals, these can be computed as \cite{Lappi:2007ku}  
\begin{eqnarray}
V_{\rm p/Pb}(\xt)=\prod_{i=1}^{N_{Y}} \exp \left( -i g \frac{\rho_{\rm p/Pb}(Y_{i},\mathbf{x})}{\boldsymbol{\nabla}_\perp^2+m^2} \right) \;,
\end{eqnarray}
for a given configuration of color charges.\footnote{We have introduced an effective mass $m$ to regulate the non-perturbative large distance behavior. If not stated otherwise we will use a fixed value of $m=0.4~{\rm GeV}$  in the following.} Here, $\boldsymbol{\nabla}_\perp^2 = \partial_i \partial^i$.
We employ a McLerran-Venugopalan type model \cite{McLerran:1993ni} for the color charge densities, which follow local Gaussian distributions
 with variance
\begin{eqnarray}\label{eq:rhorho}
g^2 \langle\rho^{a}(Y_{i},\mathbf{x})\rho^{b}(Y_{j},\mathbf{y})\rangle &=&S_{\rm p/Pb}(\mathbf{b})~\delta^{ab}~\frac{\delta_{Y_{i} Y_{j}}}{N_{Y}}~\delta^{(2)}(\mathbf{x-y})\;, \nonumber \\
\end{eqnarray}
where $\mathbf{b}=(\mathbf{x}+\mathbf{y})/2$. The spatial distributions of color charge inside the proton and the nucleus are described by $S_{\rm p/Pb}(\mathbf{b})$ and the respective models are outlined below.
\subsection{The proton}

We will consider two different models for the distribution of color charge $S_{\rm p}(\mathbf{b})$ inside the proton 
to study the effect of the spatial sub-structure of the proton on the observed correlations.

The \emph{spherical proton} model is a variant of the IP-Sat model \cite{Kowalski:2003hm} where the color charge distribution inside the proton is spherically symmetric in impact parameter space. In this case, the color charge density is proportional to the saturation scale, i.e.,
\begin{eqnarray}
\label{eq:SIPGlasma}
S_{\rm p}(\mathbf{b})=c \times Q_{s}^{2}(\sqrt{s},t(\mathbf{b}))\,.
\end{eqnarray}
$Q_s$ itself depends on the transverse position $\mathbf{b}$ via the nucleon thickness function
\begin{eqnarray}
t(\mathbf{b})=\frac{1}{2\pi B_G} \exp\left(-\frac{\mathbf{b}^{2}}{2B_G}\right).
\end{eqnarray} 
The Gaussian width is related to the (two dimensional) proton radius relevant to strong interactions as $R_{\rm p}=\sqrt{2\cdot B_G}$~\cite{Caldwell:2009ke}. Its value $B_G=4\pm 0.4~{\rm GeV}^{-2}$ as well as the dependence of $Q_{s}^{2}$  on the nucleon thickness $t(\mathbf{b})$ and the center of mass energy $\sqrt{s}$ have been extracted from fits to deep-inelastic scattering (DIS) data~\cite{Rezaeian:2012ji}.\footnote{In practice one extracts the saturation scale $Q_s(\sqrt{s},t(b))$ from the IP-Sat parametrization of the dipole scattering amplitude. More details on this procedure and the general features of this model can be found in~\cite{Schenke:2013dpa}.} We set $B_G=4\,{\rm GeV}^{-2}$ and the only free parameter in this model is the proportionality factor $c$ in Eq.~(\ref{eq:SIPGlasma}), defined as $g^4\mu^2/Q_s^2$, where $g^2\mu^2$ is the color charge square per unit transverse area. We choose to use a fixed value of $c=2$ and shall comment later on the sensitivity of our results under variation of $c$. 
    
The \emph{constituent quark proton} model was previously outlined in \cite{Schlichting:2014ipa}, where the distribution of the color charge density is concentrated around the (transverse) positions $\mathbf{x}_{\rm CQ}$ of three constituent quarks
\begin{eqnarray}
S_{p}(\mathbf{b}) = c \times \frac{3~\overline{Q}^2 }{2\pi} \sum_{n=1}^{N_{\rm CQ}} \exp \left(  -\frac{3}{2} \frac{\Big(\mathbf{b}-\mathbf{x}^{(n)}_{\rm CQ}\Big)^2}{R_{\rm CQ}^{2}} \right)\;,
\end{eqnarray}
which fluctuate from event to event according to a Gaussian distribution with expectation value $\langle\mathbf{x}^2_{\rm CQ}\rangle=B_G$. The gluon distribution around each constituent quark is spherically symmetric with a radius denoted by $R_{\rm CQ}$. We will use $R_{\rm CQ}=\sqrt{B_G}/2$  and adjust the overall strength $\overline{Q}=3~{\rm GeV}$, which yields similar results for the dipole scattering amplitude (relevant to DIS) as the spherical proton model.

\subsection{The nucleus}
Since we expect a smaller sensitivity of our results to the impact parameter dependent structure of the nucleus, we limit ourselves to a single model for the color charge distribution inside the nucleus. We first sample the positions $\mathbf{x}_{i}$ of $A=208$ (for Pb) individual nucleons according to a Wood-Saxon distribution with radius and surface parameters appropriate for a Pb nucleus. We follow the IP-Sat model and set
\begin{eqnarray}
S_{\rm Pb}(\mathbf{b})=c \times Q_{s}^{2}(\sqrt{s},T(\mathbf{b}))\;, 
\end{eqnarray}
where the thickness function $T(\mathbf{b})$ of the nucleus
\begin{eqnarray}
T(\mathbf{b})=\sum_{i=1}^{A} t(\mathbf{b}-\mathbf{x}_{i})
\end{eqnarray}
is the sum of thickness functions $t(\mathbf{b}-\mathbf{x}_{i})$ of individual nucleons. We note that for the case of a single nucleon $(A=1)$, this reduces to the spherical proton model and we employ precisely the same parametrization in both cases.

\subsection{Early-time dynamics and gluon distribution}
Solving the classical Yang-Mills equations outside the forward light-cone leads to the initial state immediately after the collision $(\tau=0^{+})$. The initial gauge fields in Fock--Schwinger gauge 
$A^\tau=(x^+ A^- + x^- A^+)/\tau=0$ are given in terms of the projectile and target fields as \cite{Kovner:1995ja,Kovner:1995ts}
\begin{align}
  \left. A^i \right |_{\tau=0^{+}} &= A^i_{\rm p} + A^i_{\rm Pb}\,,   &\left. \partial_\tau A^i \right |_{\tau=0^{+}} = 0\, ,\label{eq:init1}\\
  \left. A^\eta \right |_{\tau=0^{+}}  &= \frac{ig}{2}\left[A^i_{\rm p},A^i_{\rm Pb}\right]\,,   & \left. \partial_\tau A^\eta \right |_{\tau=0^{+}} = 0\;,\label{eq:init2}
\end{align}
and correspond to longitudinal chromo-electric and chromo magnetic fields~\cite{Lappi:2006fp}
\begin{eqnarray}
\label{eq:InitEB}
\left.E^{\eta} \right|_{\tau=0^{+}} = -2A^{\eta} \;, ~~ \left.\mathbf{E}_{\bot} \right|_{\tau=0^{+}}=0\;, ~~ \left.\mathbf{B}_{\bot} \right|_{\tau=0^{+}}=0\;, \nonumber \\
\left.B^{\eta}\right|_{\tau=0^{+}}=\partial_{x}A_{y}-\partial_{y}A_{x}-ig[A_{x},A_{y}]\;. \qquad
\end{eqnarray}
Starting from these field configurations, the early time dynamics in each event can be determined by numerically solving the classical Yang Mills equations in the forward light cone. Our numerical implementation is based on standard lattice gauge theory techniques and we refer the reader to \cite{Krasnitz:1999wc,*Krasnitz:2000gz,Lappi:2003bi,Schenke:2012fw} for more details of this procedure.

We can then extract the gluon distribution at different times of the evolution by measuring equal-time correlation functions of the gauge fields. We impose the Coulomb gauge condition $\left.\partial_{i}A^{i}\right|_{\tau}$ at the time of each measurement and follow previous works \cite{Berges:2013fga,Berges:2013eia} to compute the single particle spectrum by a projection on to transversely polarized gluon modes. The single particle distribution is then given by
\begin{eqnarray}\label{eq:singleParticleDist}
\left.\frac{dN}{d^{2}\kt dy}\right|_{\tau}=\frac{1}{(2\pi)^2} \sum_{\lambda,a} \left| \tau g^{\mu\nu} \Big( \xi_{\mu}^{\lambda,\kt*}(\tau) \overleftrightarrow{\partial_{\tau}} A_{\nu}^{a}(\tau,\kt)\Big) \right|^2 \nonumber \\
\end{eqnarray}
where  $g^{\mu\nu}=(1,-1,-1,-\tau^{-2})$ denotes the Bjorken metric, $\lambda=1,2$ labels the two transverse polarizations and $a=1,\cdots,N_{c}^{2}-1$ is the color index. In Coulomb gauge the mode functions take the form
\begin{eqnarray}
\xi_{\mu}^{(1),\kt}(\tau)&=& \frac{\sqrt{\pi}}{2|\kt|}  \begin{pmatrix} -k_y \\ k_x \\ 0 \end{pmatrix}  H^{(2)}_{0}(|\kt|\tau) \;, \label{eq:A1} \\
\xi_{\mu}^{(2),\kt}(\tau)&=&\frac{\sqrt{\pi}}{2|\kt|}  \begin{pmatrix} 0 \\ 0 \\  k_T\tau \end{pmatrix}  H^{(2)}_{1}(|\kt|\tau) \;, \label{eq:A2}
\end{eqnarray}
where $\kt=(k_{x},k_{y})$ and $H^{(2)}_{\alpha}$ denote the Hankel functions of the second type and order $\alpha$ (see \cite{Berges:2013fga} for details).

We note that the above definition of the gluon distribution is such that $dN/d^{2}\kt$ is exactly conserved for a non-interacting system. This property is important, as it will enable us to clearly distinguish between the properties of the initial state at $\tau=0^{+}$ and the effect of final state interactions at later times.

\section{Azimuthal anisotropies}
We will now discuss the measurement of azimuthal anisotropies, in particular the extraction  of $v_2(p_T)$ and $v_3(p_T)$ for gluons, using two different methods. One is based on a measurement of the single particle anisotropy in each event while the other follows more closely experimental measurements based on two-particle correlations.

\subsection{Single particle anisotropy}
Within the single particle method  we determine the Fourier coefficients $v_{2}(p_T)$ and $v_{3}(p_T)$ from the azimuthal anisotropy of the single particle spectra. Since the lattice simulation yields the single particle spectrum at discrete values of the transverse momenta $k_x$ and $k_y$, we first perform a bi-linear interpolation and divide the data into transverse momentum $|\kt|$ and azimuthal angle $\phi$ bins. While in general both the Fourier coefficients $v_{n}(p_T)$ and event plane angles $\psi_{n}(p_T)$ in Eq.~(\ref{eq:vnep}) depend on the transverse momentum $p_T$ and fluctuate event by event, we  will for the moment disregard the $p_T$ dependence of the event plane (EP) and instead compute the second and third order event plane angles $\psi_{2}(p_T^{\rm ref})$ and $\psi_{3}(p_T^{\rm ref})$ for gluons in each event as an average over a reference momentum region, which is chosen to be $1\,{\rm GeV} < p_T^{\rm ref} < 6\,{\rm GeV}$ as discussed below. We then extract the Fourier coefficients $v_{2}(EP)(p_T)$ and $v_{3}(EP)(p_T)$ with respect to the reference event plane according to
\begin{eqnarray}
v_{n}(EP)(p_T)= \frac{ \int d\phi~\frac{dN}{d^2p_T} \cos(n (\phi-\psi_{n}(p_T^{\rm ref}))}{\int d\phi~\frac{dN}{d^2 p_T} } \;.
\end{eqnarray}
Since $v_{n}(EP)(p_T)$ fluctuates from event to event our final result is obtained by performing an average of these quantities over all events.

\begin{figure*}[t!]
   \begin{center}
     \includegraphics[width=0.475\textwidth]{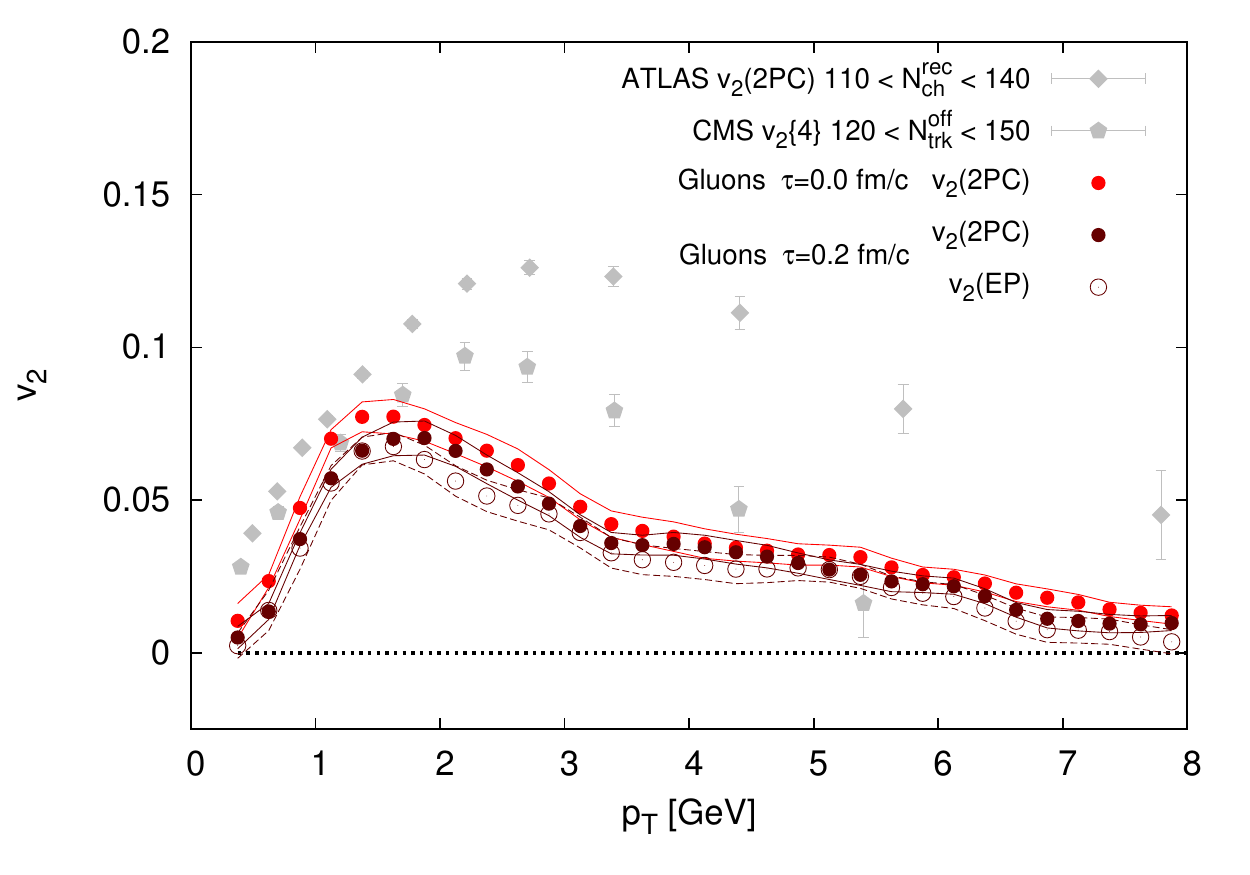}
     \includegraphics[width=0.475\textwidth]{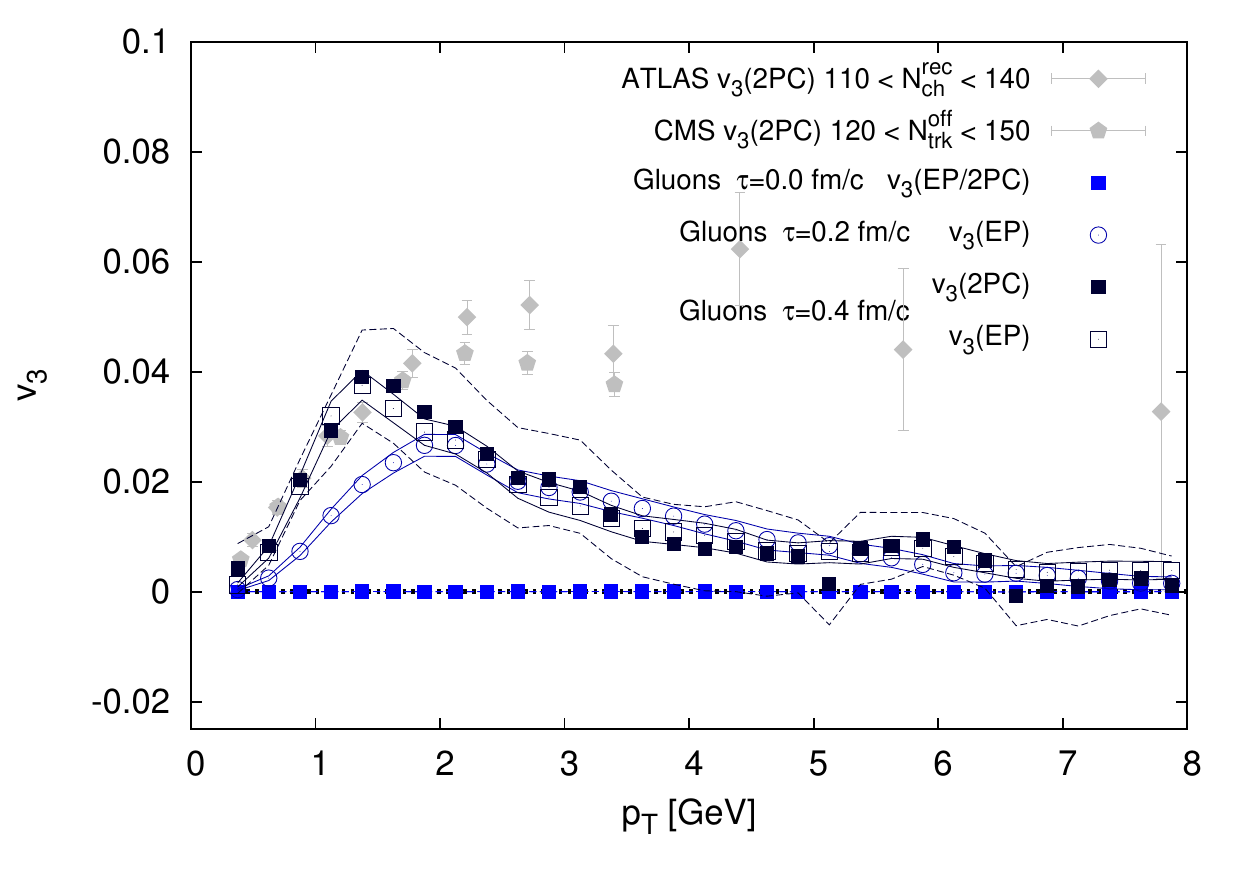}

     \caption{ \label{fig:vnEccProt} (Color online) Gluon $v_2(p_T)$ (left) and  $v_3(p_T)$ (right) at different times $\tau=0.0-0.4~{\rm fm/c}$ in p+Pb collisions at impact parameter $b=0\,{\rm fm}$ in the constituent quark proton model. Open symbols correspond to the single particle anisotropy measurement while filled symbols show the results obtained from two particle correlations. Error bands correspond to statistical errors only. Experimental results by the ATLAS~\cite{Aad:2014lta} and CMS collaboration \cite{Chatrchyan:2013nka} for inclusive hadrons are also shown as a guideline for comparison.}
      \end{center}
\end{figure*}

\subsection{Two particle correlations}
The measurement of two-particle correlations in experiments is based on the per trigger yield defined to be \cite{Chatrchyan:2013nka,Aad:2014lta}
\begin{equation}
  \frac{1}{N_{\rm trig}}\frac{dN^{\rm pair}}{d\Delta\eta d\Delta\phi} = B(0,0) \frac{S(\Delta\eta,\Delta\phi)}{B(\Delta\eta,\Delta\phi)}\, .
\end{equation}
Here $N_{\rm trig}$ is the number of trigger particles in the momentum bin under consideration,
\begin{equation}
S(\Delta\eta,\Delta\phi) =   \frac{1}{N_{\rm trig}}\frac{dN^{\rm same}}{d\Delta\eta d\Delta\phi}
\end{equation}
is the signal and 
\begin{equation}
B(\Delta\eta,\Delta\phi) =   \frac{1}{N_{\rm trig}}\frac{dN^{\rm mix}}{d\Delta\eta d\Delta\phi}
\end{equation}
is the background contribution estimated from mixed events. \\

Since there are no acceptance and efficiency corrections in the theory calculation, we can directly compute the quantity \cite{Lappi:2009xa,Kovchegov:2012nd}
\begin{align}\label{eq:dNdDeltaPhi}
 & \frac{2\pi}{N_{\rm trig}N_{\rm assoc}}\frac{dN^{{\rm pair}}}{d\Delta\phi}(k_1,k_2) = \notag\\ 
 & ~~~~ \frac{\left\langle \int_{0}^{2\pi} d\phi~~\frac{dN}{d^2k_T}(k_1,\phi)\frac{dN}{d^2k_T}(k_2,\phi+\Delta\phi) \right\rangle}
 {\frac{1}{2\pi}\int_0^{2\pi} d\Delta\phi \left\langle \int_{0}^{2\pi} d\phi~~\frac{dN}{d^2k_T}(k_1,\phi)\frac{dN}{d^2k_T}(k_2,\phi+\Delta\phi) \right\rangle }
 \,,
\end{align}
where $N_{\rm assoc}$ is the number of associated particles in the momentum bin considered and the event average over the product of single particle distributions $\langle \frac{dN}{d^2k_T^{1}}\frac{dN}{d^2k_T^{2}} \rangle$ contains the event-by-event single particle anisotropy as well as non-factorizable (connected) two-particle correlations. Note that while in experiments a rapidity gap is introduced to suppress jet-like correlations around $\Delta\eta=0$, jet-like correlations are not present in our  calculation at this order~\cite{Dusling:2012cg} and $\Delta\eta$ gaps are therefore unnecessary.

The Fourier expansion of Eq.\,(\ref{eq:dNdDeltaPhi}),
\begin{align}\label{eq:expansion}
  \frac{2\pi}{N_{\rm trig}N_{\rm assoc}}&\frac{dN^{{\rm pair}}}{d\Delta\phi}(k_1,k_2) = \notag\\ &  1 + \sum_n 2 V_{n\Delta}(k_1,k_2) \cos(n\Delta\phi)
\end{align}
yields the coefficients $V_{n\Delta}$, from which we define the two particle gluon $v_{2}(p_T)$ and $v_{3}(p_T)$ to be
\begin{equation}
  v_n(2PC)(p_T) = \frac{V_{n\Delta}(p_T,p_T^{\rm ref})}{\sqrt{V_{n\Delta}(p_T^{\rm ref},p_T^{\rm ref})}}\;,
\end{equation}
as is done by the experimental collaborations to measure $v_{n}(p_T)$ for inclusive hadrons. We choose the reference momentum range for gluons to be $1\,{\rm GeV} < p_T^{\rm ref} < 6\,{\rm GeV}$. The upper limit in $p_T$ of this range extends to somewhat larger $p_T$ than that of experimental measurements for inclusive hadrons \cite{Chatrchyan:2013nka,Aad:2014lta}.  This is to account very roughly for the fragmentation of higher momentum gluons into hadrons in the $p_T$ bin of interest. We will compute $V_{n\Delta}(p_T,p_T^{\rm ref})$ in $p_T$ bins of width $0.25\,{\rm GeV}$ defined symmetrically around the quoted $p_T$ values.  We will comment in Sec.~V on the sensitivity of our results to the reference momentum range.

\section{Numerical results for ${\text p}$+${\text P}{\text b}$ collisions}
We will now present numerical results for the azimuthal correlations of gluons in p+Pb collisions. We first study the behavior for collisions of a constituent quark proton with a lead nucleus and focus on collisions with zero impact parameter $b=0\,{\rm fm}$ without any further multiplicity selections applied. Our results for $v_2(p_T)$ and $v_3(p_T)$ of gluons -- obtained from an average over $N_{{\rm evt}}=128$ events --  are shown in the left and right panels of Fig.~\ref{fig:vnEccProt}. We also show experimental results for inclusive hadron $v_{2}(p_T)$ and $v_{3}(p_T)$ from the ATLAS~\cite{Aad:2014lta} and CMS ~\cite{Chatrchyan:2013nka} collaborations to provide an estimate of the relative magnitude and momentum dependence of the observed correlations.  We emphasize that since the gluon to hadron conversion is by no means straightforward, we do not expect a quantitative description of the data within the present framework.

We find that already at the initial time $\tau=0^{+}$ the $v_2(p_T)$ of gluons is quite large and extends up to relatively high transverse momenta, which can be seen in both the two particle correlation and single particle anisotropy measurements. The fact that both methods give rise to very similar results points to the fact that the origin of the observed $v_2(p_T)$ is due to a breaking of rotational symmetry of the single particle spectrum.
In other words, we conclude that gluons are produced with a preferred azimuthal direction in each event. 

While the emergence of a preferred azimuthal direction is well understood in the context of a collective expansion of the system, we emphasize that the non-zero gluon $v_2(p_T)$ at the initial time can not be attributed to collective flow effects. As a consequence of the initial color fields in Eq.~(\ref{eq:InitEB}) the energy momentum tensor at $\tau=0^{+}$ is strictly diagonal $T_{\mu\nu}(\xt)={\rm diag}(\epsilon(\xt),\epsilon(\xt),\epsilon(\xt),-\epsilon(\xt))$  and the Poynting vector $\vec{E} \times \vec{B}$ characterizing the energy momentum flow vanishes identically. Instead, the observed $v_{2}(p_T)$ should be attributed to anisotropic gluon production from the fluctuating color fields inside the projectile and target. 
 
Our result for $v_2$ and $v_3$ from two-particle correlations can be separated into three contributions. First, the Glasma graph contribution, which corresponds to the connected graphs analyzed in \cite{Dumitru:2008wn,Lappi:2009xa,Dumitru:2010iy,Dusling:2012iga,Dusling:2012cg,Dusling:2012wy,Kovchegov:2012nd,Dusling:2013oia}. Second, a contribution from disconnected graphs, which takes into account event-by-event breaking of rotational symmetry similar to the effect discussed in \cite{Kovner:2012jm,Dumitru:2014dra,Dumitru:2014vka,Lappi:2015vha,Dumitru:2014yza}, and finally, a contribution from final state interactions represented by the Yang-Mills dynamics of the produced gluon fields. 

The contribution from disconnected graphs can be understood in the following simplified picture: One can imagine the gluon production process as the partons inside the projectile scattering off a high-energy nucleus. Each parton experiences the color electric field inside the nucleus whereby it receives a momentum kick in a certain direction. In each event the color electric field of the nucleus is characterized by several domains with different orientations in azimuthal direction and color space. While partons scattering off different domains receive a different kick, partons (with the same color charge) scattering off the same domain will receive a kick in the same azimuthal direction. Hence, depending on the number of uncorrelated domains probed by the projectile, the initial state particle production can be significantly anisotropic.

The Glasma graph contribution directly probes the color structure within each domain and corresponds to genuine non-factorizable two-particle correlations. Depending on the degree of polarization of a single domain, the two contributions can be of comparable size \cite{Dumitru:2014yza}. 

\begin{figure*}[t!]
   \begin{center}
     \includegraphics[width=0.475\textwidth]{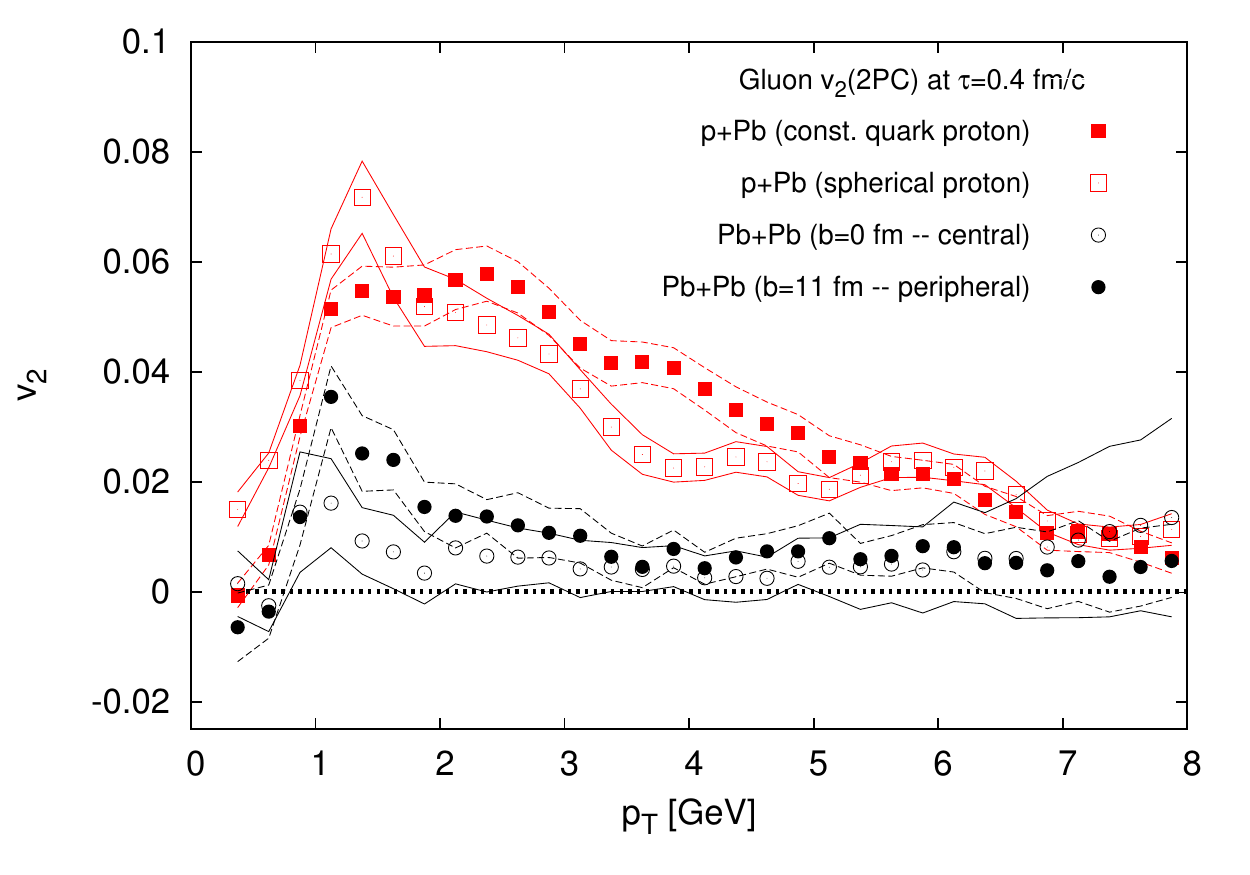}
     \includegraphics[width=0.475\textwidth]{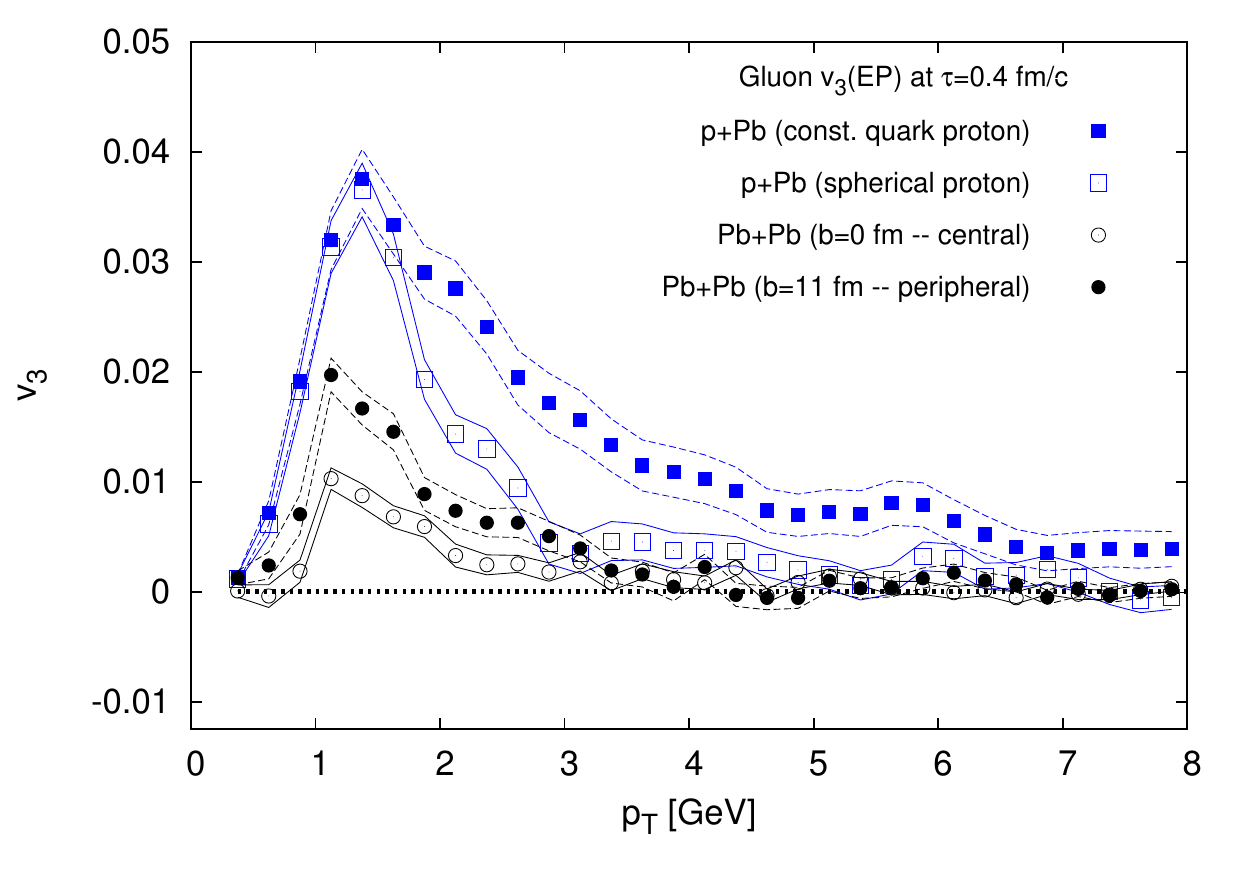}

     \caption{ \label{fig:vnComparison} (Color online) Dependence of gluon $v_2(p_T)$ (left) and  $v_3(p_T)$ (right)  at time $\tau=0.4~{\rm fm/c}$ on the collision geometry and system size. Shown are results for p+Pb collisions with two different models for the proton structure and Pb+Pb collisions at two different impact parameters.}
   \end{center}
\end{figure*}

With regard to the effect of final state interactions on the observed gluon $v_{2}(p_T)$, we find that the classical Yang-Mills evolution leads to slight decrease of the observed $v_{2}(p_T)$ with time. This decrease may be attributed to gluon rescattering after the collision. The overall effect of the classical Yang-Mills evolution on $v_{2}$ is very small and the initial state value provides a very good estimate for the correlation at later times. We note again that this is conceptually quite different from a hydrodynamic mechanism, where $v_{2}(p_T)$ is gradually build up as a function of time. 

We now turn to a discussion of the gluon $v_{3}(p_T)$ shown in the right panel of Fig.~\ref{fig:vnEccProt} for which the interpretation is drastically different. We find that the initial state $v_{3}(p_T)$ vanishes identically at $\tau=0^{+}$. This is a consequence of the vanishing transverse color electric $(\mathbf{E}_{\bot})$ and color magnetic fields $(\mathbf{B}_{\bot})$ at the initial time. With only the longitudinal color electric $(E^{\eta})$ and color magnetic  $(B^{\eta})$ fields being non-zero at $\tau=0^{+}$, the gluon spectrum in Eq.~(\ref{eq:singleParticleDist}) takes the form
\begin{eqnarray}
\label{eq:SpectrumT0}
\left.\frac{dN}{d^{2}\kt dy}\right|_{\tau=0^{+}}&=&\frac{1}{(2\pi)^2} \sum_{a=1}^{N_{c}^{2}-1} \left[ \frac{B^{\eta}_{a}(\kt)B^{\eta}_{a}(-\kt)}{\pi|\kt|^2} \right. \nonumber \\
&& \qquad \quad  \left.+\frac{E^{\eta}_{a}(\kt)E^{\eta}_{a}(-\kt)}{\pi|\kt|^2}\right]\;.
\end{eqnarray}
The above spectrum is manifestly symmetric under $\mathbf{k}_T \leftrightarrow -\mathbf{k}_T$ -- hence odd Fourier harmonics vanish at the initial time. As a result, we can attribute the observation of a non-zero $v_{3}(p_T)$ in our framework at later times exclusively to coherent final state effects included via the classical Yang-Mills evolution. 

Quantitatively, we find that already at very early times $\tau\sim0.2~{\rm fm/c}$ the classical Yang-Mills evolution leads to the build up of a sizable gluon $v_{3}(p_T)$  extending to relatively large transverse momenta. Beyond $\tau=0.2~{\rm fm/c}$, the $v_{3}$ of hard gluons remains approximately constant while the low momentum $v_{3}$ continues to show an increase with time. When we follow the classical Yang-Mills dynamics to even later times the system becomes more and more dilute and approaches a free streaming behavior around $\tau=0.4~{\rm fm}/c$ as previously reported in \cite{Schenke:2012fw}.

We note the agreement between the two particle correlation and single particle anisotropy measurement of $v_{3}(p_T)$ again points to a correlation between many particles in each event. While it may appear suggestive that the build up of energy-momentum flow may cause a non-zero $v_{3}(p_T)$ at later times, {\it to our surprise, we did not observe a significant correlation between the global initial state eccentricity $\varepsilon_3$ and the $p_T$ integrated momentum space anisotropy $v_{3}$ on an event-by-event basis}. However, it is possible that there is a correlation between the final $v_3$ and triangular anisotropies on shorter length scales than the ones probed by the global $\varepsilon_3$ - such geometrical features are however difficult to extract.
Generally, a simple description of the non-linear dynamics underlying the emergence of a non-zero $v_{3}(p_T)$ remains elusive - a deeper understanding of this effect is clearly desirable.

\section{Sensitivity to proton structure and collision geometry} 
We will now study the effect of the collision geometry and system size on the azimuthal correlations. Our results are compactly summarized in Fig.~\ref{fig:vnComparison} which shows a comparison of the gluon $v_{2}(p_T)$ and $v_{3}(p_T)$ at time $\tau=0.4~{\rm fm/c}$ after the collision. 

We first analyze the effect of a more detailed substructure of the proton projectile on the correlations in p+Pb collisions by comparing a spherical proton with one composed of three valence quarks. Generally, a finer substructure leads to larger $v_2$ and $v_3$ at transverse momenta $p_T\gtrsim 2\,{\rm GeV}$ -- corresponding to wave lengths on size scales much smaller than the nucleon size.  However the overall effect of the proton's geometry for the observed azimuthal anisotropy of gluons is far less significant than one would expect in a mechanism that generates azimuthal anisotropies via final state collective effects \cite{Schenke:2014zha}. Since the origin of the observed correlations is due to the microscopic structure of color fields, one instead expects the correlations to be approximately independent of the global event geometry. Our result in Fig.~\ref{fig:vnComparison} confirms this picture. 

We have also considered variations of the non-perturbative mass scale $m$ and the coefficient $c$ in Eq.~(\ref{eq:SIPGlasma}) by a factor of two. While in both cases we did not observe a significant effect on the overall magnitude of the observed correlations, we found that for smaller (larger) values of $m$ the correlations extend over a slightly larger (smaller) transverse momentum range. We note further that changing the reference momentum range $p_T^{\rm ref}$ can also have a significant effect on the transverse momentum dependence of the signal which is somewhat more pronounced for $v_{2}(p_T)$ as compared to $v_{3}(p_T)$. While the gluon spectrum is generally anisotropic up to high momenta, the event plane angles at different transverse momenta $\psi_{n}(p_T)$ are not strongly correlated across the entire range of momenta. Considering a larger (smaller) reference momentum range can therefore lead to a slower (faster) decrease at high momenta. Similar observations have recently been reported in a related study~\cite{Lappi:2015vha} in which the proton was treated as a dilute object.
 
We conclude our study with a comparison between proton-nucleus (p+Pb) and nucleus-nucleus (Pb+Pb) collisions. In the latter case, we have analyzed collisions at two different impact parameters corresponding to central ($b=0~{\rm fm}$) and peripheral ($b=11~{\rm fm}$) collision.\footnote{When classified in terms of centrality percentile these correspond approximately to the $0-5\%$ and respectively $50-60\%$ centrality classes.} 

We find that in both central and peripheral nucleus-nucleus collisions the correlations between gluons are much smaller compared to p+Pb collisions. Qualitatively this difference can be understood when considering the different number of localized domains of fluctuating color fields responsible for the production of gluons in the different collision systems. While in proton-lead collisions, particles are produced from a small number of different domains inside the nucleus, the larger overlap area in lead-lead collisions gives rise to particle production from a much larger number of different domains. Since different domains are uncorrelated with each other the azimuthal anisotropy of the gluon spectrum decreases with the number of domains (see also \cite{Dumitru:2014dra}). Consequently, the initial state  momentum space anisotropy is much smaller in Pb+Pb collisions as compared to p+Pb collisions.

\section{Conclusions}
We have presented results for the azimuthal anisotropy of the single and double inclusive gluon distributions in p+Pb and Pb+Pb collisions obtained from classical Yang-Mills simulations. Both the proton and the nucleus have been treated as dense QCD objects with high gluon occupancy. This description is appropriate for the early time space-time evolution of high multiplicity p+Pb as well as heavy ion collisions at high energies.

Gluons produced in p+Pb collisions show a significant $v_2(p_T)$ already at the initial time immediately after the collision. Further evolution governed by the Yang-Mills equations modifies this $v_2(p_T)$ only slightly. In contrast, odd harmonics of gluons are initially exactly zero, but significant values of $v_3(p_T)$ are built up within times $\tau \lesssim 0.4\,{\rm fm}$ of the classical Yang-Mills evolution. These momentum space anisotropies at early times are uncorrelated with the \emph{global} spatial anisotropy, in contrast to anisotropies generated by collective flow. 

 Our results indicate that in p+Pb collisions there are significant contributions to $v_2$ and $v_3$ from the initial production (in the case of $v_2$ alone) and the early time non-equilibrium dynamics within the first half fermi of evolution.  These effects cannot be neglected and any calculation based merely on final state effects is thus incomplete.
 
A similar analysis of Pb+Pb collisions reveals a different picture. Initial and early-time contributions to $v_2$ and $v_3$ are small, indicating that for larger systems final state collective effects are indeed the dominant mechanism for generating the observed anisotropies -- at least at momenta $p_T\leq 2$ GeV, where the presence of such effects is very plausible. The difference between p+Pb and Pb+Pb collisions can be understood as a consequence, in this framework, of  anisotropies being generated due to localized domains of color fields.  A large number of mutually uncorrelated domains probed in Pb+Pb collisions leads to a nearly isotropic gluon spectrum. Hence initial state contributions to $v_2$ and $v_3$ are small for large collision systems.

While our study provides a first attempt to quantify the importance of initial state effects in high-multiplicity proton-nucleus collisions, we expect systematic comparisons of p+A collisions with deuteron-nucleus (d+A) \cite{Adare:2013piz,Wang:2014qiw,LiYiforSTAR:2014mea,Abdelwahab:2014cvd} and $^3$He+Au \cite{he3:2014is} collisions at RHIC to provide further insight into the relative significance of initial state and final state effects in small systems. Forthcoming p+A collisions at RHIC will also help clarify the role of nucleon fluctuations relative to sub-nucleon scale effects in small systems besides providing a comparison of results for identical systems at vastly different center of mass energies.  

\section*{Acknowledgments}
We thank Adrian Dumitru, Kevin Dusling, and Yuri Kovchegov for useful discussions. BPS, SS, and RV are supported under DOE Contract No. DE-SC0012704.
This research used resources of the National Energy Research Scientific Computing Center, which is supported by the Office of Science of the U.S. Department of Energy under Contract No. DE-AC02-05CH11231. SS gratefully acknowledges a Goldhaber Distinguished Fellowship from Brookhaven Science Associates. BPS is supported by a DOE Office of Science Early Career Award.

\vspace{-0.5cm}
\bibliography{spires}

\begin{thebibliography}{67}%
\makeatletter
\providecommand \@ifxundefined [1]{%
 \@ifx{#1\undefined}
}%
\providecommand \@ifnum [1]{%
 \ifnum #1\expandafter \@firstoftwo
 \else \expandafter \@secondoftwo
 \fi
}%
\providecommand \@ifx [1]{%
 \ifx #1\expandafter \@firstoftwo
 \else \expandafter \@secondoftwo
 \fi
}%
\providecommand \natexlab [1]{#1}%
\providecommand \enquote  [1]{``#1''}%
\providecommand \bibnamefont  [1]{#1}%
\providecommand \bibfnamefont [1]{#1}%
\providecommand \citenamefont [1]{#1}%
\providecommand \href@noop [0]{\@secondoftwo}%
\providecommand \href [0]{\begingroup \@sanitize@url \@href}%
\providecommand \@href[1]{\@@startlink{#1}\@@href}%
\providecommand \@@href[1]{\endgroup#1\@@endlink}%
\providecommand \@sanitize@url [0]{\catcode `\\12\catcode `\$12\catcode
  `\&12\catcode `\#12\catcode `\^12\catcode `\_12\catcode `\%12\relax}%
\providecommand \@@startlink[1]{}%
\providecommand \@@endlink[0]{}%
\providecommand \url  [0]{\begingroup\@sanitize@url \@url }%
\providecommand \@url [1]{\endgroup\@href {#1}{\urlprefix }}%
\providecommand \urlprefix  [0]{URL }%
\providecommand \Eprint [0]{\href }%
\providecommand \doibase [0]{http://dx.doi.org/}%
\providecommand \selectlanguage [0]{\@gobble}%
\providecommand \bibinfo  [0]{\@secondoftwo}%
\providecommand \bibfield  [0]{\@secondoftwo}%
\providecommand \translation [1]{[#1]}%
\providecommand \BibitemOpen [0]{}%
\providecommand \bibitemStop [0]{}%
\providecommand \bibitemNoStop [0]{.\EOS\space}%
\providecommand \EOS [0]{\spacefactor3000\relax}%
\providecommand \BibitemShut  [1]{\csname bibitem#1\endcsname}%
\let\auto@bib@innerbib\@empty
%</preamble>
\bibitem [{\citenamefont {Khachatryan}\ \emph {et~al.}(2010)\citenamefont
  {Khachatryan} \emph {et~al.}}]{Khachatryan:2010gv}%
  \BibitemOpen
  \bibfield  {author} {\bibinfo {author} {\bibfnamefont {V.}~\bibnamefont
  {Khachatryan}} \emph {et~al.} (\bibinfo {collaboration} {CMS
  Collaboration}),\ }\href {\doibase 10.1007/JHEP09(2010)091} {\bibfield
  {journal} {\bibinfo  {journal} {JHEP}\ }\textbf {\bibinfo {volume} {1009}},\
  \bibinfo {pages} {091} (\bibinfo {year} {2010})},\ \Eprint
  {http://arxiv.org/abs/1009.4122} {arXiv:1009.4122 [hep-ex]} \BibitemShut
  {NoStop}%
%%CITATION = 1009.4122;%%
\bibitem [{\citenamefont {Chatrchyan}\ \emph
  {et~al.}(2013{\natexlab{a}})\citenamefont {Chatrchyan} \emph
  {et~al.}}]{CMS:2012qk}%
  \BibitemOpen
  \bibfield  {author} {\bibinfo {author} {\bibfnamefont {S.}~\bibnamefont
  {Chatrchyan}} \emph {et~al.} (\bibinfo {collaboration} {CMS Collaboration}),\
  }\href {\doibase 10.1016/j.physletb.2012.11.025} {\bibfield  {journal}
  {\bibinfo  {journal} {Phys.Lett.}\ }\textbf {\bibinfo {volume} {B718}},\
  \bibinfo {pages} {795} (\bibinfo {year} {2013}{\natexlab{a}})},\ \Eprint
  {http://arxiv.org/abs/1210.5482} {arXiv:1210.5482 [nucl-ex]} \BibitemShut
  {NoStop}%
%%CITATION = ARXIV:1210.5482;%%
\bibitem [{\citenamefont {Chatrchyan}\ \emph
  {et~al.}(2013{\natexlab{b}})\citenamefont {Chatrchyan} \emph
  {et~al.}}]{Chatrchyan:2013nka}%
  \BibitemOpen
  \bibfield  {author} {\bibinfo {author} {\bibfnamefont {S.}~\bibnamefont
  {Chatrchyan}} \emph {et~al.} (\bibinfo {collaboration} {CMS Collaboration}),\
  }\href {\doibase 10.1016/j.physletb.2013.06.028} {\bibfield  {journal}
  {\bibinfo  {journal} {Phys.Lett.}\ }\textbf {\bibinfo {volume} {B724}},\
  \bibinfo {pages} {213} (\bibinfo {year} {2013}{\natexlab{b}})},\ \Eprint
  {http://arxiv.org/abs/1305.0609} {arXiv:1305.0609 [nucl-ex]} \BibitemShut
  {NoStop}%
%%CITATION = ARXIV:1305.0609;%%
\bibitem [{\citenamefont {Aad}\ \emph {et~al.}(2012{\natexlab{a}})\citenamefont
  {Aad} \emph {et~al.}}]{Aad:2012gla}%
  \BibitemOpen
  \bibfield  {author} {\bibinfo {author} {\bibfnamefont {G.}~\bibnamefont
  {Aad}} \emph {et~al.} (\bibinfo {collaboration} {ATLAS Collaboration}),\
  }\href@noop {} {\  (\bibinfo {year} {2012}{\natexlab{a}})},\ \Eprint
  {http://arxiv.org/abs/1212.5198} {arXiv:1212.5198 [hep-ex]} \BibitemShut
  {NoStop}%
%%CITATION = ARXIV:1212.5198;%%
\bibitem [{\citenamefont {Aad}\ \emph {et~al.}(2013)\citenamefont {Aad} \emph
  {et~al.}}]{Aad:2013fja}%
  \BibitemOpen
  \bibfield  {author} {\bibinfo {author} {\bibfnamefont {G.}~\bibnamefont
  {Aad}} \emph {et~al.} (\bibinfo {collaboration} {ATLAS Collaboration}),\
  }\href@noop {} {\  (\bibinfo {year} {2013})},\ \Eprint
  {http://arxiv.org/abs/1303.2084} {arXiv:1303.2084 [hep-ex]} \BibitemShut
  {NoStop}%
%%CITATION = ARXIV:1303.2084;%%
\bibitem [{\citenamefont {Aad}\ \emph {et~al.}(2014)\citenamefont {Aad} \emph
  {et~al.}}]{Aad:2014lta}%
  \BibitemOpen
  \bibfield  {author} {\bibinfo {author} {\bibfnamefont {G.}~\bibnamefont
  {Aad}} \emph {et~al.} (\bibinfo {collaboration} {ATLAS Collaboration}),\
  }\href {\doibase 10.1103/PhysRevC.90.044906} {\bibfield  {journal} {\bibinfo
  {journal} {Phys.Rev.}\ }\textbf {\bibinfo {volume} {C90}},\ \bibinfo {pages}
  {044906} (\bibinfo {year} {2014})},\ \Eprint {http://arxiv.org/abs/1409.1792}
  {arXiv:1409.1792 [hep-ex]} \BibitemShut {NoStop}%
%%CITATION = ARXIV:1409.1792;%%
\bibitem [{\citenamefont {Abelev}\ \emph
  {et~al.}(2013{\natexlab{a}})\citenamefont {Abelev} \emph
  {et~al.}}]{Abelev:2012ola}%
  \BibitemOpen
  \bibfield  {author} {\bibinfo {author} {\bibfnamefont {B.}~\bibnamefont
  {Abelev}} \emph {et~al.} (\bibinfo {collaboration} {ALICE Collaboration}),\
  }\href {\doibase 10.1016/j.physletb.2013.01.012} {\bibfield  {journal}
  {\bibinfo  {journal} {Phys.Lett.}\ }\textbf {\bibinfo {volume} {B719}},\
  \bibinfo {pages} {29} (\bibinfo {year} {2013}{\natexlab{a}})},\ \Eprint
  {http://arxiv.org/abs/1212.2001} {arXiv:1212.2001 [nucl-ex]} \BibitemShut
  {NoStop}%
%%CITATION = ARXIV:1212.2001;%%
\bibitem [{\citenamefont {Abelev}\ \emph
  {et~al.}(2013{\natexlab{b}})\citenamefont {Abelev} \emph
  {et~al.}}]{ABELEV:2013wsa}%
  \BibitemOpen
  \bibfield  {author} {\bibinfo {author} {\bibfnamefont {B.~B.}\ \bibnamefont
  {Abelev}} \emph {et~al.} (\bibinfo {collaboration} {ALICE Collaboration}),\
  }\href {\doibase 10.1016/j.physletb.2013.08.024} {\bibfield  {journal}
  {\bibinfo  {journal} {Phys.Lett.}\ }\textbf {\bibinfo {volume} {B726}},\
  \bibinfo {pages} {164} (\bibinfo {year} {2013}{\natexlab{b}})},\ \Eprint
  {http://arxiv.org/abs/1307.3237} {arXiv:1307.3237 [nucl-ex]} \BibitemShut
  {NoStop}%
%%CITATION = ARXIV:1307.3237;%%
\bibitem [{\citenamefont {Chatrchyan}\ \emph {et~al.}(2011)\citenamefont
  {Chatrchyan} \emph {et~al.}}]{Chatrchyan:2011eka}%
  \BibitemOpen
  \bibfield  {author} {\bibinfo {author} {\bibfnamefont {S.}~\bibnamefont
  {Chatrchyan}} \emph {et~al.} (\bibinfo {collaboration} {CMS Collaboration}),\
  }\href {\doibase 10.1007/JHEP07(2011)076} {\bibfield  {journal} {\bibinfo
  {journal} {JHEP}\ }\textbf {\bibinfo {volume} {1107}},\ \bibinfo {pages}
  {076} (\bibinfo {year} {2011})},\ \Eprint {http://arxiv.org/abs/1105.2438}
  {arXiv:1105.2438 [nucl-ex]} \BibitemShut {NoStop}%
%%CITATION = ARXIV:1105.2438;%%
\bibitem [{\citenamefont {Aamodt}\ \emph {et~al.}(2011)\citenamefont {Aamodt}
  \emph {et~al.}}]{ALICE:2011ab}%
  \BibitemOpen
  \bibfield  {author} {\bibinfo {author} {\bibfnamefont {K.}~\bibnamefont
  {Aamodt}} \emph {et~al.} (\bibinfo {collaboration} {ALICE Collaboration}),\
  }\href {\doibase 10.1103/PhysRevLett.107.032301} {\bibfield  {journal}
  {\bibinfo  {journal} {Phys.Rev.Lett.}\ }\textbf {\bibinfo {volume} {107}},\
  \bibinfo {pages} {032301} (\bibinfo {year} {2011})}\BibitemShut {NoStop}%
%%CITATION = ARXIV:1105.3865;%%
\bibitem [{\citenamefont {Aad}\ \emph {et~al.}(2012{\natexlab{b}})\citenamefont
  {Aad} \emph {et~al.}}]{ATLAS:2012at}%
  \BibitemOpen
  \bibfield  {author} {\bibinfo {author} {\bibfnamefont {G.}~\bibnamefont
  {Aad}} \emph {et~al.} (\bibinfo {collaboration} {ATLAS Collaboration}),\
  }\href {\doibase 10.1103/PhysRevC.86.014907} {\bibfield  {journal} {\bibinfo
  {journal} {Phys.Rev.}\ }\textbf {\bibinfo {volume} {C86}},\ \bibinfo {pages}
  {014907} (\bibinfo {year} {2012}{\natexlab{b}})}\BibitemShut {NoStop}%
%%CITATION = ARXIV:1203.3087;%%
\bibitem [{\citenamefont {Gale}\ \emph {et~al.}(2013)\citenamefont {Gale},
  \citenamefont {Jeon}, \citenamefont {Schenke}, \citenamefont {Tribedy},\ and\
  \citenamefont {Venugopalan}}]{Gale:2012rq}%
  \BibitemOpen
  \bibfield  {author} {\bibinfo {author} {\bibfnamefont {C.}~\bibnamefont
  {Gale}}, \bibinfo {author} {\bibfnamefont {S.}~\bibnamefont {Jeon}}, \bibinfo
  {author} {\bibfnamefont {B.}~\bibnamefont {Schenke}}, \bibinfo {author}
  {\bibfnamefont {P.}~\bibnamefont {Tribedy}}, \ and\ \bibinfo {author}
  {\bibfnamefont {R.}~\bibnamefont {Venugopalan}},\ }\href {\doibase
  10.1103/PhysRevLett.110.012302} {\bibfield  {journal} {\bibinfo  {journal}
  {Phys.Rev.Lett.}\ }\textbf {\bibinfo {volume} {110}},\ \bibinfo {pages}
  {012302} (\bibinfo {year} {2013})},\ \Eprint {http://arxiv.org/abs/1209.6330}
  {arXiv:1209.6330 [nucl-th]} \BibitemShut {NoStop}%
%%CITATION = ARXIV:1209.6330;%%
\bibitem [{\citenamefont {Bozek}\ and\ \citenamefont
  {Broniowski}(2013)}]{Bozek:2013uha}%
  \BibitemOpen
  \bibfield  {author} {\bibinfo {author} {\bibfnamefont {P.}~\bibnamefont
  {Bozek}}\ and\ \bibinfo {author} {\bibfnamefont {W.}~\bibnamefont
  {Broniowski}},\ }\href {\doibase 10.1103/PhysRevC.88.014903} {\bibfield
  {journal} {\bibinfo  {journal} {Phys.Rev.}\ }\textbf {\bibinfo {volume}
  {C88}},\ \bibinfo {pages} {014903} (\bibinfo {year} {2013})},\ \Eprint
  {http://arxiv.org/abs/1304.3044} {arXiv:1304.3044 [nucl-th]} \BibitemShut
  {NoStop}%
%%CITATION = ARXIV:1304.3044;%%
\bibitem [{\citenamefont {Bozek}\ \emph {et~al.}(2013)\citenamefont {Bozek},
  \citenamefont {Broniowski},\ and\ \citenamefont {Torrieri}}]{Bozek:2013ska}%
  \BibitemOpen
  \bibfield  {author} {\bibinfo {author} {\bibfnamefont {P.}~\bibnamefont
  {Bozek}}, \bibinfo {author} {\bibfnamefont {W.}~\bibnamefont {Broniowski}}, \
  and\ \bibinfo {author} {\bibfnamefont {G.}~\bibnamefont {Torrieri}},\ }\href
  {\doibase 10.1103/PhysRevLett.111.172303} {\bibfield  {journal} {\bibinfo
  {journal} {Phys.Rev.Lett.}\ }\textbf {\bibinfo {volume} {111}},\ \bibinfo
  {pages} {172303} (\bibinfo {year} {2013})}\BibitemShut {NoStop}%
%%CITATION = ARXIV:1307.5060;%%
\bibitem [{\citenamefont {Bzdak}\ and\ \citenamefont
  {Ma}(2014)}]{Bzdak:2014dia}%
  \BibitemOpen
  \bibfield  {author} {\bibinfo {author} {\bibfnamefont {A.}~\bibnamefont
  {Bzdak}}\ and\ \bibinfo {author} {\bibfnamefont {G.-L.}\ \bibnamefont {Ma}},\
  }\href {\doibase 10.1103/PhysRevLett.113.252301} {\bibfield  {journal}
  {\bibinfo  {journal} {Phys.Rev.Lett.}\ }\textbf {\bibinfo {volume} {113}},\
  \bibinfo {pages} {252301} (\bibinfo {year} {2014})},\ \Eprint
  {http://arxiv.org/abs/1406.2804} {arXiv:1406.2804 [hep-ph]} \BibitemShut
  {NoStop}%
%%CITATION = ARXIV:1406.2804;%%
\bibitem [{\citenamefont {Qin}\ and\ \citenamefont
  {Muller}(2014)}]{Qin:2013bha}%
  \BibitemOpen
  \bibfield  {author} {\bibinfo {author} {\bibfnamefont {G.-Y.}\ \bibnamefont
  {Qin}}\ and\ \bibinfo {author} {\bibfnamefont {B.}~\bibnamefont {Muller}},\
  }\href {\doibase 10.1103/PhysRevC.89.044902} {\bibfield  {journal} {\bibinfo
  {journal} {Phys.Rev.}\ }\textbf {\bibinfo {volume} {C89}},\ \bibinfo {pages}
  {044902} (\bibinfo {year} {2014})},\ \Eprint {http://arxiv.org/abs/1306.3439}
  {arXiv:1306.3439 [nucl-th]} \BibitemShut {NoStop}%
%%CITATION = ARXIV:1306.3439;%%
\bibitem [{\citenamefont {Werner}\ \emph {et~al.}(2014)\citenamefont {Werner},
  \citenamefont {Bleicher}, \citenamefont {Guiot}, \citenamefont {Karpenko},\
  and\ \citenamefont {Pierog}}]{Werner:2013ipa}%
  \BibitemOpen
  \bibfield  {author} {\bibinfo {author} {\bibfnamefont {K.}~\bibnamefont
  {Werner}}, \bibinfo {author} {\bibfnamefont {M.}~\bibnamefont {Bleicher}},
  \bibinfo {author} {\bibfnamefont {B.}~\bibnamefont {Guiot}}, \bibinfo
  {author} {\bibfnamefont {I.}~\bibnamefont {Karpenko}}, \ and\ \bibinfo
  {author} {\bibfnamefont {T.}~\bibnamefont {Pierog}},\ }\href {\doibase
  10.1103/PhysRevLett.112.232301} {\bibfield  {journal} {\bibinfo  {journal}
  {Phys.Rev.Lett.}\ }\textbf {\bibinfo {volume} {112}},\ \bibinfo {pages}
  {232301} (\bibinfo {year} {2014})},\ \Eprint {http://arxiv.org/abs/1307.4379}
  {arXiv:1307.4379 [nucl-th]} \BibitemShut {NoStop}%
%%CITATION = ARXIV:1307.4379;%%
\bibitem [{\citenamefont {Bzdak}\ \emph {et~al.}(2013)\citenamefont {Bzdak},
  \citenamefont {Schenke}, \citenamefont {Tribedy},\ and\ \citenamefont
  {Venugopalan}}]{Bzdak:2013zma}%
  \BibitemOpen
  \bibfield  {author} {\bibinfo {author} {\bibfnamefont {A.}~\bibnamefont
  {Bzdak}}, \bibinfo {author} {\bibfnamefont {B.}~\bibnamefont {Schenke}},
  \bibinfo {author} {\bibfnamefont {P.}~\bibnamefont {Tribedy}}, \ and\
  \bibinfo {author} {\bibfnamefont {R.}~\bibnamefont {Venugopalan}},\
  }\href@noop {} {\  (\bibinfo {year} {2013})},\ \Eprint
  {http://arxiv.org/abs/1304.3403} {arXiv:1304.3403 [nucl-th]} \BibitemShut
  {NoStop}%
%%CITATION = ARXIV:1304.3403;%%
\bibitem [{\citenamefont {Niemi}\ and\ \citenamefont
  {Denicol}(2014)}]{Niemi:2014wta}%
  \BibitemOpen
  \bibfield  {author} {\bibinfo {author} {\bibfnamefont {H.}~\bibnamefont
  {Niemi}}\ and\ \bibinfo {author} {\bibfnamefont {G.}~\bibnamefont
  {Denicol}},\ }\href@noop {} {\  (\bibinfo {year} {2014})},\ \Eprint
  {http://arxiv.org/abs/1404.7327} {arXiv:1404.7327 [nucl-th]} \BibitemShut
  {NoStop}%
%%CITATION = ARXIV:1404.7327;%%
\bibitem [{\citenamefont {Schenke}\ and\ \citenamefont
  {Venugopalan}(2014)}]{Schenke:2014zha}%
  \BibitemOpen
  \bibfield  {author} {\bibinfo {author} {\bibfnamefont {B.}~\bibnamefont
  {Schenke}}\ and\ \bibinfo {author} {\bibfnamefont {R.}~\bibnamefont
  {Venugopalan}},\ }\href {\doibase 10.1103/PhysRevLett.113.102301} {\bibfield
  {journal} {\bibinfo  {journal} {Phys.Rev.Lett.}\ }\textbf {\bibinfo {volume}
  {113}},\ \bibinfo {pages} {102301} (\bibinfo {year} {2014})},\ \Eprint
  {http://arxiv.org/abs/1405.3605} {arXiv:1405.3605 [nucl-th]} \BibitemShut
  {NoStop}%
%%CITATION = ARXIV:1405.3605;%%
\bibitem [{\citenamefont {Dumitru}\ \emph {et~al.}(2011)\citenamefont
  {Dumitru}, \citenamefont {Dusling}, \citenamefont {Gelis}, \citenamefont
  {Jalilian-Marian}, \citenamefont {Lappi},\ and\ \citenamefont
  {Venugopalan}}]{Dumitru:2010iy}%
  \BibitemOpen
  \bibfield  {author} {\bibinfo {author} {\bibfnamefont {A.}~\bibnamefont
  {Dumitru}}, \bibinfo {author} {\bibfnamefont {K.}~\bibnamefont {Dusling}},
  \bibinfo {author} {\bibfnamefont {F.}~\bibnamefont {Gelis}}, \bibinfo
  {author} {\bibfnamefont {J.}~\bibnamefont {Jalilian-Marian}}, \bibinfo
  {author} {\bibfnamefont {T.}~\bibnamefont {Lappi}}, \ and\ \bibinfo {author}
  {\bibfnamefont {R.}~\bibnamefont {Venugopalan}},\ }\href {\doibase
  10.1016/j.physletb.2011.01.024} {\bibfield  {journal} {\bibinfo  {journal}
  {Phys. Lett.}\ }\textbf {\bibinfo {volume} {B697}},\ \bibinfo {pages} {21}
  (\bibinfo {year} {2011})},\ \Eprint {http://arxiv.org/abs/1009.5295}
  {arXiv:1009.5295 [hep-ph]} \BibitemShut {NoStop}%
\bibitem [{\citenamefont {Dusling}\ and\ \citenamefont
  {Venugopalan}(2012{\natexlab{a}})}]{Dusling:2012iga}%
  \BibitemOpen
  \bibfield  {author} {\bibinfo {author} {\bibfnamefont {K.}~\bibnamefont
  {Dusling}}\ and\ \bibinfo {author} {\bibfnamefont {R.}~\bibnamefont
  {Venugopalan}},\ }\href {\doibase 10.1103/PhysRevLett.108.262001} {\bibfield
  {journal} {\bibinfo  {journal} {Phys.Rev.Lett.}\ }\textbf {\bibinfo {volume}
  {108}},\ \bibinfo {pages} {262001} (\bibinfo {year} {2012}{\natexlab{a}})},\
  \Eprint {http://arxiv.org/abs/1201.2658} {arXiv:1201.2658 [hep-ph]}
  \BibitemShut {NoStop}%
%%CITATION = ARXIV:1201.2658;%%
\bibitem [{\citenamefont {Dusling}\ and\ \citenamefont
  {Venugopalan}(2012{\natexlab{b}})}]{Dusling:2012cg}%
  \BibitemOpen
  \bibfield  {author} {\bibinfo {author} {\bibfnamefont {K.}~\bibnamefont
  {Dusling}}\ and\ \bibinfo {author} {\bibfnamefont {R.}~\bibnamefont
  {Venugopalan}},\ }\href@noop {} {\  (\bibinfo {year} {2012}{\natexlab{b}})},\
  \Eprint {http://arxiv.org/abs/1210.3890} {arXiv:1210.3890 [hep-ph]}
  \BibitemShut {NoStop}%
%%CITATION = ARXIV:1210.3890;%%
\bibitem [{\citenamefont {Dusling}\ and\ \citenamefont
  {Venugopalan}(2012{\natexlab{c}})}]{Dusling:2012wy}%
  \BibitemOpen
  \bibfield  {author} {\bibinfo {author} {\bibfnamefont {K.}~\bibnamefont
  {Dusling}}\ and\ \bibinfo {author} {\bibfnamefont {R.}~\bibnamefont
  {Venugopalan}},\ }\href@noop {} {\  (\bibinfo {year} {2012}{\natexlab{c}})},\
  \Eprint {http://arxiv.org/abs/1211.3701} {arXiv:1211.3701 [hep-ph]}
  \BibitemShut {NoStop}%
%%CITATION = ARXIV:1211.3701;%%
\bibitem [{\citenamefont {Dusling}\ and\ \citenamefont
  {Venugopalan}(2013)}]{Dusling:2013oia}%
  \BibitemOpen
  \bibfield  {author} {\bibinfo {author} {\bibfnamefont {K.}~\bibnamefont
  {Dusling}}\ and\ \bibinfo {author} {\bibfnamefont {R.}~\bibnamefont
  {Venugopalan}},\ }\href@noop {} {\  (\bibinfo {year} {2013})},\ \Eprint
  {http://arxiv.org/abs/1302.7018} {arXiv:1302.7018 [hep-ph]} \BibitemShut
  {NoStop}%
%%CITATION = ARXIV:1302.7018;%%
\bibitem [{\citenamefont {Dusling}\ and\ \citenamefont
  {Venugopalan}(2014)}]{Dusling:2014oha}%
  \BibitemOpen
  \bibfield  {author} {\bibinfo {author} {\bibfnamefont {K.}~\bibnamefont
  {Dusling}}\ and\ \bibinfo {author} {\bibfnamefont {R.}~\bibnamefont
  {Venugopalan}},\ }\href {\doibase 10.1016/j.nuclphysa.2014.09.024} {\bibfield
   {journal} {\bibinfo  {journal} {Nucl.Phys.}\ }\textbf {\bibinfo {volume}
  {A931}},\ \bibinfo {pages} {283} (\bibinfo {year} {2014})}\BibitemShut
  {NoStop}%
%%CITATION = NUPHA,A931,283;%%
\bibitem [{\citenamefont {Kovner}\ and\ \citenamefont
  {Lublinsky}(2013)}]{Kovner:2012jm}%
  \BibitemOpen
  \bibfield  {author} {\bibinfo {author} {\bibfnamefont {A.}~\bibnamefont
  {Kovner}}\ and\ \bibinfo {author} {\bibfnamefont {M.}~\bibnamefont
  {Lublinsky}},\ }\href {\doibase 10.1142/S0218301313300014} {\bibfield
  {journal} {\bibinfo  {journal} {Int.J.Mod.Phys.}\ }\textbf {\bibinfo {volume}
  {E22}},\ \bibinfo {pages} {1330001} (\bibinfo {year} {2013})},\ \Eprint
  {http://arxiv.org/abs/1211.1928} {arXiv:1211.1928 [hep-ph]} \BibitemShut
  {NoStop}%
%%CITATION = ARXIV:1211.1928;%%
\bibitem [{\citenamefont {Dumitru}\ and\ \citenamefont
  {Giannini}(2014)}]{Dumitru:2014dra}%
  \BibitemOpen
  \bibfield  {author} {\bibinfo {author} {\bibfnamefont {A.}~\bibnamefont
  {Dumitru}}\ and\ \bibinfo {author} {\bibfnamefont {A.~V.}\ \bibnamefont
  {Giannini}},\ }\href@noop {} {\  (\bibinfo {year} {2014})},\ \Eprint
  {http://arxiv.org/abs/1406.5781} {arXiv:1406.5781 [hep-ph]} \BibitemShut
  {NoStop}%
%%CITATION = ARXIV:1406.5781;%%
\bibitem [{\citenamefont {Dumitru}\ and\ \citenamefont
  {Skokov}(2014)}]{Dumitru:2014vka}%
  \BibitemOpen
  \bibfield  {author} {\bibinfo {author} {\bibfnamefont {A.}~\bibnamefont
  {Dumitru}}\ and\ \bibinfo {author} {\bibfnamefont {V.}~\bibnamefont
  {Skokov}},\ }\href@noop {} {\  (\bibinfo {year} {2014})},\ \Eprint
  {http://arxiv.org/abs/1411.6630} {arXiv:1411.6630 [hep-ph]} \BibitemShut
  {NoStop}%
%%CITATION = ARXIV:1411.6630;%%
\bibitem [{\citenamefont {Noronha}\ and\ \citenamefont
  {Dumitru}(2014)}]{Noronha:2014vva}%
  \BibitemOpen
  \bibfield  {author} {\bibinfo {author} {\bibfnamefont {J.}~\bibnamefont
  {Noronha}}\ and\ \bibinfo {author} {\bibfnamefont {A.}~\bibnamefont
  {Dumitru}},\ }\href {\doibase 10.1103/PhysRevD.89.094008} {\bibfield
  {journal} {\bibinfo  {journal} {Phys.Rev.}\ }\textbf {\bibinfo {volume}
  {D89}},\ \bibinfo {pages} {094008} (\bibinfo {year} {2014})},\ \Eprint
  {http://arxiv.org/abs/1401.4467} {arXiv:1401.4467 [hep-ph]} \BibitemShut
  {NoStop}%
%%CITATION = ARXIV:1401.4467;%%
\bibitem [{\citenamefont {Gyulassy}\ \emph {et~al.}(2014)\citenamefont
  {Gyulassy}, \citenamefont {Levai}, \citenamefont {Vitev},\ and\ \citenamefont
  {Biro}}]{Gyulassy:2014cfa}%
  \BibitemOpen
  \bibfield  {author} {\bibinfo {author} {\bibfnamefont {M.}~\bibnamefont
  {Gyulassy}}, \bibinfo {author} {\bibfnamefont {P.}~\bibnamefont {Levai}},
  \bibinfo {author} {\bibfnamefont {I.}~\bibnamefont {Vitev}}, \ and\ \bibinfo
  {author} {\bibfnamefont {T.}~\bibnamefont {Biro}},\ }\href {\doibase
  10.1103/PhysRevD.90.054025} {\bibfield  {journal} {\bibinfo  {journal}
  {Phys.Rev.}\ }\textbf {\bibinfo {volume} {D90}},\ \bibinfo {pages} {054025}
  (\bibinfo {year} {2014})},\ \Eprint {http://arxiv.org/abs/1405.7825}
  {arXiv:1405.7825 [hep-ph]} \BibitemShut {NoStop}%
%%CITATION = ARXIV:1405.7825;%%
\bibitem [{\citenamefont {Lappi}(2015)}]{Lappi:2015vha}%
  \BibitemOpen
  \bibfield  {author} {\bibinfo {author} {\bibfnamefont {T.}~\bibnamefont
  {Lappi}},\ }\href@noop {} {\  (\bibinfo {year} {2015})},\ \Eprint
  {http://arxiv.org/abs/1501.05505} {arXiv:1501.05505 [hep-ph]} \BibitemShut
  {NoStop}%
%%CITATION = ARXIV:1501.05505;%%
\bibitem [{\citenamefont {Krasnitz}\ and\ \citenamefont
  {Venugopalan}(2000)}]{Krasnitz:1999wc}%
  \BibitemOpen
  \bibfield  {author} {\bibinfo {author} {\bibfnamefont {A.}~\bibnamefont
  {Krasnitz}}\ and\ \bibinfo {author} {\bibfnamefont {R.}~\bibnamefont
  {Venugopalan}},\ }\href@noop {} {\bibfield  {journal} {\bibinfo  {journal}
  {Phys. Rev. Lett.}\ }\textbf {\bibinfo {volume} {84}},\ \bibinfo {pages}
  {4309} (\bibinfo {year} {2000})}\BibitemShut {NoStop}%
%%CITATION = HEP-PH/9909203;%%
\bibitem [{\citenamefont {Krasnitz}\ and\ \citenamefont
  {Venugopalan}(2001)}]{Krasnitz:2000gz}%
  \BibitemOpen
  \bibfield  {author} {\bibinfo {author} {\bibfnamefont {A.}~\bibnamefont
  {Krasnitz}}\ and\ \bibinfo {author} {\bibfnamefont {R.}~\bibnamefont
  {Venugopalan}},\ }\href {\doibase 10.1103/PhysRevLett. 86.1717} {\bibfield
  {journal} {\bibinfo  {journal} {Phys. Rev. Lett.}\ }\textbf {\bibinfo
  {volume} {86}},\ \bibinfo {pages} {1717} (\bibinfo {year}
  {2001})}\BibitemShut {NoStop}%
%%CITATION = HEP-PH/0007108;%%
\bibitem [{\citenamefont {Lappi}(2003)}]{Lappi:2003bi}%
  \BibitemOpen
  \bibfield  {author} {\bibinfo {author} {\bibfnamefont {T.}~\bibnamefont
  {Lappi}},\ }\href {\doibase 10.1103/PhysRevC.67.054903} {\bibfield  {journal}
  {\bibinfo  {journal} {Phys. Rev.}\ }\textbf {\bibinfo {volume} {C67}},\
  \bibinfo {pages} {054903} (\bibinfo {year} {2003})}\BibitemShut {NoStop}%
%%CITATION = HEP-PH/0303076;%%
\bibitem [{\citenamefont {Krasnitz}\ \emph
  {et~al.}(2003{\natexlab{a}})\citenamefont {Krasnitz}, \citenamefont {Nara},\
  and\ \citenamefont {Venugopalan}}]{Krasnitz:2002ng}%
  \BibitemOpen
  \bibfield  {author} {\bibinfo {author} {\bibfnamefont {A.}~\bibnamefont
  {Krasnitz}}, \bibinfo {author} {\bibfnamefont {Y.}~\bibnamefont {Nara}}, \
  and\ \bibinfo {author} {\bibfnamefont {R.}~\bibnamefont {Venugopalan}},\
  }\href@noop {} {\bibfield  {journal} {\bibinfo  {journal} {Phys. Lett.}\
  }\textbf {\bibinfo {volume} {B554}},\ \bibinfo {pages} {21} (\bibinfo {year}
  {2003}{\natexlab{a}})}\BibitemShut {NoStop}%
%%CITATION = HEP-PH/0204361;%%
\bibitem [{\citenamefont {Krasnitz}\ \emph
  {et~al.}(2003{\natexlab{b}})\citenamefont {Krasnitz}, \citenamefont {Nara},\
  and\ \citenamefont {Venugopalan}}]{Krasnitz:2002mn}%
  \BibitemOpen
  \bibfield  {author} {\bibinfo {author} {\bibfnamefont {A.}~\bibnamefont
  {Krasnitz}}, \bibinfo {author} {\bibfnamefont {Y.}~\bibnamefont {Nara}}, \
  and\ \bibinfo {author} {\bibfnamefont {R.}~\bibnamefont {Venugopalan}},\
  }\href@noop {} {\bibfield  {journal} {\bibinfo  {journal} {Nucl. Phys.}\
  }\textbf {\bibinfo {volume} {A717}},\ \bibinfo {pages} {268} (\bibinfo {year}
  {2003}{\natexlab{b}})}\BibitemShut {NoStop}%
%%CITATION = HEP-PH/0209269;%%
\bibitem [{\citenamefont {Schenke}\ \emph {et~al.}(2014)\citenamefont
  {Schenke}, \citenamefont {Tribedy},\ and\ \citenamefont
  {Venugopalan}}]{Schenke:2013dpa}%
  \BibitemOpen
  \bibfield  {author} {\bibinfo {author} {\bibfnamefont {B.}~\bibnamefont
  {Schenke}}, \bibinfo {author} {\bibfnamefont {P.}~\bibnamefont {Tribedy}}, \
  and\ \bibinfo {author} {\bibfnamefont {R.}~\bibnamefont {Venugopalan}},\
  }\href {\doibase 10.1103/PhysRevC.89.024901} {\bibfield  {journal} {\bibinfo
  {journal} {Phys.Rev.}\ }\textbf {\bibinfo {volume} {C89}},\ \bibinfo {pages}
  {024901} (\bibinfo {year} {2014})},\ \Eprint {http://arxiv.org/abs/1311.3636}
  {arXiv:1311.3636 [hep-ph]} \BibitemShut {NoStop}%
%%CITATION = ARXIV:1311.3636;%%
\bibitem [{\citenamefont {Schenke}\ \emph
  {et~al.}(2012{\natexlab{a}})\citenamefont {Schenke}, \citenamefont
  {Tribedy},\ and\ \citenamefont {Venugopalan}}]{Schenke:2012wb}%
  \BibitemOpen
  \bibfield  {author} {\bibinfo {author} {\bibfnamefont {B.}~\bibnamefont
  {Schenke}}, \bibinfo {author} {\bibfnamefont {P.}~\bibnamefont {Tribedy}}, \
  and\ \bibinfo {author} {\bibfnamefont {R.}~\bibnamefont {Venugopalan}},\
  }\href@noop {} {\bibfield  {journal} {\bibinfo  {journal} {Phys. Rev. Lett.}\
  }\textbf {\bibinfo {volume} {108}},\ \bibinfo {pages} {252301} (\bibinfo
  {year} {2012}{\natexlab{a}})}\BibitemShut {NoStop}%
%%CITATION = ARXIV:1202.6646;%%
\bibitem [{\citenamefont {Schenke}\ \emph
  {et~al.}(2012{\natexlab{b}})\citenamefont {Schenke}, \citenamefont
  {Tribedy},\ and\ \citenamefont {Venugopalan}}]{Schenke:2012fw}%
  \BibitemOpen
  \bibfield  {author} {\bibinfo {author} {\bibfnamefont {B.}~\bibnamefont
  {Schenke}}, \bibinfo {author} {\bibfnamefont {P.}~\bibnamefont {Tribedy}}, \
  and\ \bibinfo {author} {\bibfnamefont {R.}~\bibnamefont {Venugopalan}},\
  }\href {\doibase 10.1103/PhysRevC.86.034908} {\bibfield  {journal} {\bibinfo
  {journal} {Phys.Rev.}\ }\textbf {\bibinfo {volume} {C86}},\ \bibinfo {pages}
  {034908} (\bibinfo {year} {2012}{\natexlab{b}})},\ \Eprint
  {http://arxiv.org/abs/1206.6805} {arXiv:1206.6805 [hep-ph]} \BibitemShut
  {NoStop}%
%%CITATION = ARXIV:1206.6805;%%
\bibitem [{\citenamefont {Schlichting}\ and\ \citenamefont
  {Schenke}(2014)}]{Schlichting:2014ipa}%
  \BibitemOpen
  \bibfield  {author} {\bibinfo {author} {\bibfnamefont {S.}~\bibnamefont
  {Schlichting}}\ and\ \bibinfo {author} {\bibfnamefont {B.}~\bibnamefont
  {Schenke}},\ }\href {\doibase 10.1016/j.physletb.2014.10.068} {\bibfield
  {journal} {\bibinfo  {journal} {Phys.Lett.}\ }\textbf {\bibinfo {volume}
  {B739}},\ \bibinfo {pages} {313} (\bibinfo {year} {2014})},\ \Eprint
  {http://arxiv.org/abs/1407.8458} {arXiv:1407.8458 [hep-ph]} \BibitemShut
  {NoStop}%
%%CITATION = ARXIV:1407.8458;%%
\bibitem [{\citenamefont {Iancu}\ \emph {et~al.}(2002)\citenamefont {Iancu},
  \citenamefont {Leonidov},\ and\ \citenamefont {McLerran}}]{Iancu:2002xk}%
  \BibitemOpen
  \bibfield  {author} {\bibinfo {author} {\bibfnamefont {E.}~\bibnamefont
  {Iancu}}, \bibinfo {author} {\bibfnamefont {A.}~\bibnamefont {Leonidov}}, \
  and\ \bibinfo {author} {\bibfnamefont {L.}~\bibnamefont {McLerran}},\ }in\
  \href@noop {} {\emph {\bibinfo {booktitle} {Cargese 2001, QCD perspectives on
  hot and dense matter}}}\ (\bibinfo {year} {2002})\ pp.\ \bibinfo {pages}
  {74--145},\ \Eprint {http://arxiv.org/abs/hep-ph/0202270} {hep-ph/0202270}
  \BibitemShut {NoStop}%
%%CITATION = HEP-PH 0202270;%%
\bibitem [{\citenamefont {Iancu}\ and\ \citenamefont
  {Venugopalan}(2003)}]{Iancu:2003xm}%
  \BibitemOpen
  \bibfield  {author} {\bibinfo {author} {\bibfnamefont {E.}~\bibnamefont
  {Iancu}}\ and\ \bibinfo {author} {\bibfnamefont {R.}~\bibnamefont
  {Venugopalan}},\ }in\ \href@noop {} {\emph {\bibinfo {booktitle} {Quark gluon
  plasma}}},\ \bibinfo {editor} {edited by\ \bibinfo {editor} {\bibfnamefont
  {R.}~\bibnamefont {Hwa}}\ and\ \bibinfo {editor} {\bibfnamefont {X.~N.}\
  \bibnamefont {Wang}}}\ (\bibinfo  {publisher} {World Scientific},\ \bibinfo
  {year} {2003})\ \Eprint {http://arxiv.org/abs/hep-ph/0303204}
  {hep-ph/0303204} \BibitemShut {NoStop}%
%%CITATION = HEP-PH 0303204;%%
\bibitem [{\citenamefont {Gelis}\ \emph {et~al.}(2010)\citenamefont {Gelis},
  \citenamefont {Iancu}, \citenamefont {Jalilian-Marian},\ and\ \citenamefont
  {Venugopalan}}]{Gelis:2010nm}%
  \BibitemOpen
  \bibfield  {author} {\bibinfo {author} {\bibfnamefont {F.}~\bibnamefont
  {Gelis}}, \bibinfo {author} {\bibfnamefont {E.}~\bibnamefont {Iancu}},
  \bibinfo {author} {\bibfnamefont {J.}~\bibnamefont {Jalilian-Marian}}, \ and\
  \bibinfo {author} {\bibfnamefont {R.}~\bibnamefont {Venugopalan}},\ }\href
  {\doibase 10.1146/annurev.nucl.010909.083629} {\bibfield  {journal} {\bibinfo
   {journal} {Ann.Rev.Nucl.Part.Sci.}\ }\textbf {\bibinfo {volume} {60}},\
  \bibinfo {pages} {463} (\bibinfo {year} {2010})}\BibitemShut {NoStop}%
\bibitem [{\citenamefont {McLerran}\ and\ \citenamefont
  {Venugopalan}(1994{\natexlab{a}})}]{McLerran:1993ni}%
  \BibitemOpen
  \bibfield  {author} {\bibinfo {author} {\bibfnamefont {L.~D.}\ \bibnamefont
  {McLerran}}\ and\ \bibinfo {author} {\bibfnamefont {R.}~\bibnamefont
  {Venugopalan}},\ }\href {\doibase 10.1103/PhysRevD.49.2233} {\bibfield
  {journal} {\bibinfo  {journal} {Phys. Rev.}\ }\textbf {\bibinfo {volume}
  {D49}},\ \bibinfo {pages} {2233} (\bibinfo {year}
  {1994}{\natexlab{a}})}\BibitemShut {NoStop}%
%%CITATION = HEP-PH/9309289;%%
\bibitem [{\citenamefont {McLerran}\ and\ \citenamefont
  {Venugopalan}(1994{\natexlab{b}})}]{McLerran:1993ka}%
  \BibitemOpen
  \bibfield  {author} {\bibinfo {author} {\bibfnamefont {L.~D.}\ \bibnamefont
  {McLerran}}\ and\ \bibinfo {author} {\bibfnamefont {R.}~\bibnamefont
  {Venugopalan}},\ }\href {\doibase 10.1103/PhysRevD.49.3352} {\bibfield
  {journal} {\bibinfo  {journal} {Phys. Rev.}\ }\textbf {\bibinfo {volume}
  {D49}},\ \bibinfo {pages} {3352} (\bibinfo {year}
  {1994}{\natexlab{b}})}\BibitemShut {NoStop}%
%%CITATION = HEP-PH/9311205;%%
\bibitem [{\citenamefont {McLerran}\ and\ \citenamefont
  {Venugopalan}(1994{\natexlab{c}})}]{McLerran:1994vd}%
  \BibitemOpen
  \bibfield  {author} {\bibinfo {author} {\bibfnamefont {L.~D.}\ \bibnamefont
  {McLerran}}\ and\ \bibinfo {author} {\bibfnamefont {R.}~\bibnamefont
  {Venugopalan}},\ }\href {\doibase 10.1103/PhysRevD.50.2225} {\bibfield
  {journal} {\bibinfo  {journal} {Phys. Rev.}\ }\textbf {\bibinfo {volume}
  {D50}},\ \bibinfo {pages} {2225} (\bibinfo {year}
  {1994}{\natexlab{c}})}\BibitemShut {NoStop}%
%%CITATION = HEP-PH/9402335;%%
\bibitem [{\citenamefont {Jalilian-Marian}\ \emph {et~al.}(1997)\citenamefont
  {Jalilian-Marian}, \citenamefont {Kovner}, \citenamefont {McLerran},\ and\
  \citenamefont {Weigert}}]{JalilianMarian:1996xn}%
  \BibitemOpen
  \bibfield  {author} {\bibinfo {author} {\bibfnamefont {J.}~\bibnamefont
  {Jalilian-Marian}}, \bibinfo {author} {\bibfnamefont {A.}~\bibnamefont
  {Kovner}}, \bibinfo {author} {\bibfnamefont {L.~D.}\ \bibnamefont
  {McLerran}}, \ and\ \bibinfo {author} {\bibfnamefont {H.}~\bibnamefont
  {Weigert}},\ }\href {\doibase 10.1103/PhysRevD.55.5414} {\bibfield  {journal}
  {\bibinfo  {journal} {Phys. Rev.}\ }\textbf {\bibinfo {volume} {D55}},\
  \bibinfo {pages} {5414} (\bibinfo {year} {1997})}\BibitemShut {NoStop}%
%%CITATION = HEP-PH/9606337;%%
\bibitem [{\citenamefont {Kovchegov}(1996)}]{Kovchegov:1996ty}%
  \BibitemOpen
  \bibfield  {author} {\bibinfo {author} {\bibfnamefont {Y.~V.}\ \bibnamefont
  {Kovchegov}},\ }\href {\doibase 10.1103/PhysRevD.54.5463} {\bibfield
  {journal} {\bibinfo  {journal} {Phys. Rev.}\ }\textbf {\bibinfo {volume}
  {D54}},\ \bibinfo {pages} {5463} (\bibinfo {year} {1996})}\BibitemShut
  {NoStop}%
%%CITATION = HEP-PH/9605446;%%
\bibitem [{\citenamefont {Lappi}(2008)}]{Lappi:2007ku}%
  \BibitemOpen
  \bibfield  {author} {\bibinfo {author} {\bibfnamefont {T.}~\bibnamefont
  {Lappi}},\ }\href {\doibase 10.1140/epjc/s10052-008-0588-4} {\bibfield
  {journal} {\bibinfo  {journal} {Eur. Phys. J.}\ }\textbf {\bibinfo {volume}
  {C55}},\ \bibinfo {pages} {285} (\bibinfo {year} {2008})}\BibitemShut
  {NoStop}%
%%CITATION = 0711.3039;%%
\bibitem [{\citenamefont {Kowalski}\ and\ \citenamefont
  {Teaney}(2003)}]{Kowalski:2003hm}%
  \BibitemOpen
  \bibfield  {author} {\bibinfo {author} {\bibfnamefont {H.}~\bibnamefont
  {Kowalski}}\ and\ \bibinfo {author} {\bibfnamefont {D.}~\bibnamefont
  {Teaney}},\ }\href {\doibase 10.1103/PhysRevD.68.114005} {\bibfield
  {journal} {\bibinfo  {journal} {Phys. Rev.}\ }\textbf {\bibinfo {volume}
  {D68}},\ \bibinfo {pages} {114005} (\bibinfo {year} {2003})}\BibitemShut
  {NoStop}%
%%CITATION = HEP-PH/0304189;%%
\bibitem [{\citenamefont {Caldwell}\ and\ \citenamefont
  {Kowalski}(2010)}]{Caldwell:2009ke}%
  \BibitemOpen
  \bibfield  {author} {\bibinfo {author} {\bibfnamefont {A.}~\bibnamefont
  {Caldwell}}\ and\ \bibinfo {author} {\bibfnamefont {H.}~\bibnamefont
  {Kowalski}},\ }\href {\doibase 10.1103/PhysRevC.81.025203} {\bibfield
  {journal} {\bibinfo  {journal} {Phys. Rev.}\ }\textbf {\bibinfo {volume}
  {C81}},\ \bibinfo {pages} {025203} (\bibinfo {year} {2010})},\ \Eprint
  {http://arxiv.org/abs/0909.1254} {arXiv:0909.1254 [hep-ph]} \BibitemShut
  {NoStop}%
%%CITATION = PHRVA,C81,025203;%%
\bibitem [{\citenamefont {Rezaeian}\ \emph {et~al.}(2013)\citenamefont
  {Rezaeian}, \citenamefont {Siddikov}, \citenamefont {Van~de Klundert},\ and\
  \citenamefont {Venugopalan}}]{Rezaeian:2012ji}%
  \BibitemOpen
  \bibfield  {author} {\bibinfo {author} {\bibfnamefont {A.~H.}\ \bibnamefont
  {Rezaeian}}, \bibinfo {author} {\bibfnamefont {M.}~\bibnamefont {Siddikov}},
  \bibinfo {author} {\bibfnamefont {M.}~\bibnamefont {Van~de Klundert}}, \ and\
  \bibinfo {author} {\bibfnamefont {R.}~\bibnamefont {Venugopalan}},\
  }\href@noop {} {\bibfield  {journal} {\bibinfo  {journal} {Phys.Rev.}\
  }\textbf {\bibinfo {volume} {D87}},\ \bibinfo {pages} {034002} (\bibinfo
  {year} {2013})},\ \Eprint {http://arxiv.org/abs/1212.2974} {arXiv:1212.2974
  [hep-ph]} \BibitemShut {NoStop}%
%%CITATION = ARXIV:1212.2974;%%
\bibitem [{\citenamefont {Kovner}\ \emph
  {et~al.}(1995{\natexlab{a}})\citenamefont {Kovner}, \citenamefont
  {McLerran},\ and\ \citenamefont {Weigert}}]{Kovner:1995ja}%
  \BibitemOpen
  \bibfield  {author} {\bibinfo {author} {\bibfnamefont {A.}~\bibnamefont
  {Kovner}}, \bibinfo {author} {\bibfnamefont {L.~D.}\ \bibnamefont
  {McLerran}}, \ and\ \bibinfo {author} {\bibfnamefont {H.}~\bibnamefont
  {Weigert}},\ }\href {\doibase 10.1103/PhysRevD.52.6231} {\bibfield  {journal}
  {\bibinfo  {journal} {Phys. Rev.}\ }\textbf {\bibinfo {volume} {D52}},\
  \bibinfo {pages} {6231} (\bibinfo {year} {1995}{\natexlab{a}})}\BibitemShut
  {NoStop}%
%%CITATION = HEP-PH/9502289;%%
\bibitem [{\citenamefont {Kovner}\ \emph
  {et~al.}(1995{\natexlab{b}})\citenamefont {Kovner}, \citenamefont
  {McLerran},\ and\ \citenamefont {Weigert}}]{Kovner:1995ts}%
  \BibitemOpen
  \bibfield  {author} {\bibinfo {author} {\bibfnamefont {A.}~\bibnamefont
  {Kovner}}, \bibinfo {author} {\bibfnamefont {L.~D.}\ \bibnamefont
  {McLerran}}, \ and\ \bibinfo {author} {\bibfnamefont {H.}~\bibnamefont
  {Weigert}},\ }\href {\doibase 10.1103/PhysRevD.52.3809} {\bibfield  {journal}
  {\bibinfo  {journal} {Phys. Rev.}\ }\textbf {\bibinfo {volume} {D52}},\
  \bibinfo {pages} {3809} (\bibinfo {year} {1995}{\natexlab{b}})}\BibitemShut
  {NoStop}%
%%CITATION = HEP-PH/9505320;%%
\bibitem [{\citenamefont {Lappi}\ and\ \citenamefont
  {McLerran}(2006)}]{Lappi:2006fp}%
  \BibitemOpen
  \bibfield  {author} {\bibinfo {author} {\bibfnamefont {T.}~\bibnamefont
  {Lappi}}\ and\ \bibinfo {author} {\bibfnamefont {L.}~\bibnamefont
  {McLerran}},\ }\href {\doibase 10.1016/j.nuclphysa.2006.04.001} {\bibfield
  {journal} {\bibinfo  {journal} {Nucl. Phys.}\ }\textbf {\bibinfo {volume}
  {A772}},\ \bibinfo {pages} {200} (\bibinfo {year} {2006})},\ \Eprint
  {http://arxiv.org/abs/hep-ph/0602189} {arXiv:hep-ph/0602189} \BibitemShut
  {NoStop}%
%%CITATION = HEP-PH/0602189;%%
\bibitem [{\citenamefont {Berges}\ \emph
  {et~al.}(2014{\natexlab{a}})\citenamefont {Berges}, \citenamefont
  {Boguslavski}, \citenamefont {Schlichting},\ and\ \citenamefont
  {Venugopalan}}]{Berges:2013fga}%
  \BibitemOpen
  \bibfield  {author} {\bibinfo {author} {\bibfnamefont {J.}~\bibnamefont
  {Berges}}, \bibinfo {author} {\bibfnamefont {K.}~\bibnamefont {Boguslavski}},
  \bibinfo {author} {\bibfnamefont {S.}~\bibnamefont {Schlichting}}, \ and\
  \bibinfo {author} {\bibfnamefont {R.}~\bibnamefont {Venugopalan}},\ }\href
  {\doibase 10.1103/PhysRevD.89.114007} {\bibfield  {journal} {\bibinfo
  {journal} {Phys.Rev.}\ }\textbf {\bibinfo {volume} {D89}},\ \bibinfo {pages}
  {114007} (\bibinfo {year} {2014}{\natexlab{a}})},\ \Eprint
  {http://arxiv.org/abs/1311.3005} {arXiv:1311.3005 [hep-ph]} \BibitemShut
  {NoStop}%
%%CITATION = ARXIV:1311.3005;%%
\bibitem [{\citenamefont {Berges}\ \emph
  {et~al.}(2014{\natexlab{b}})\citenamefont {Berges}, \citenamefont
  {Boguslavski}, \citenamefont {Schlichting},\ and\ \citenamefont
  {Venugopalan}}]{Berges:2013eia}%
  \BibitemOpen
  \bibfield  {author} {\bibinfo {author} {\bibfnamefont {J.}~\bibnamefont
  {Berges}}, \bibinfo {author} {\bibfnamefont {K.}~\bibnamefont {Boguslavski}},
  \bibinfo {author} {\bibfnamefont {S.}~\bibnamefont {Schlichting}}, \ and\
  \bibinfo {author} {\bibfnamefont {R.}~\bibnamefont {Venugopalan}},\ }\href
  {\doibase 10.1103/PhysRevD.89.074011} {\bibfield  {journal} {\bibinfo
  {journal} {Phys.Rev.}\ }\textbf {\bibinfo {volume} {D89}},\ \bibinfo {pages}
  {074011} (\bibinfo {year} {2014}{\natexlab{b}})},\ \Eprint
  {http://arxiv.org/abs/1303.5650} {arXiv:1303.5650 [hep-ph]} \BibitemShut
  {NoStop}%
%%CITATION = ARXIV:1303.5650;%%
\bibitem [{\citenamefont {Lappi}\ \emph {et~al.}(2010)\citenamefont {Lappi},
  \citenamefont {Srednyak},\ and\ \citenamefont {Venugopalan}}]{Lappi:2009xa}%
  \BibitemOpen
  \bibfield  {author} {\bibinfo {author} {\bibfnamefont {T.}~\bibnamefont
  {Lappi}}, \bibinfo {author} {\bibfnamefont {S.}~\bibnamefont {Srednyak}}, \
  and\ \bibinfo {author} {\bibfnamefont {R.}~\bibnamefont {Venugopalan}},\
  }\href {\doibase 10.1007/JHEP01(2010)066} {\bibfield  {journal} {\bibinfo
  {journal} {JHEP}\ }\textbf {\bibinfo {volume} {01}},\ \bibinfo {pages} {066}
  (\bibinfo {year} {2010})}\BibitemShut {NoStop}%
%%CITATION = 0911.2068;%%
\bibitem [{\citenamefont {Kovchegov}\ and\ \citenamefont
  {Wertepny}(2013)}]{Kovchegov:2012nd}%
  \BibitemOpen
  \bibfield  {author} {\bibinfo {author} {\bibfnamefont {Y.~V.}\ \bibnamefont
  {Kovchegov}}\ and\ \bibinfo {author} {\bibfnamefont {D.~E.}\ \bibnamefont
  {Wertepny}},\ }\href {\doibase 10.1016/j.nuclphysa.2013.03.006} {\bibfield
  {journal} {\bibinfo  {journal} {Nucl.Phys.}\ }\textbf {\bibinfo {volume}
  {A906}},\ \bibinfo {pages} {50} (\bibinfo {year} {2013})},\ \Eprint
  {http://arxiv.org/abs/1212.1195} {arXiv:1212.1195} \BibitemShut {NoStop}%
%%CITATION = ARXIV:1212.1195;%%
\bibitem [{\citenamefont {Dumitru}\ \emph {et~al.}(2008)\citenamefont
  {Dumitru}, \citenamefont {Gelis}, \citenamefont {McLerran},\ and\
  \citenamefont {Venugopalan}}]{Dumitru:2008wn}%
  \BibitemOpen
  \bibfield  {author} {\bibinfo {author} {\bibfnamefont {A.}~\bibnamefont
  {Dumitru}}, \bibinfo {author} {\bibfnamefont {F.}~\bibnamefont {Gelis}},
  \bibinfo {author} {\bibfnamefont {L.}~\bibnamefont {McLerran}}, \ and\
  \bibinfo {author} {\bibfnamefont {R.}~\bibnamefont {Venugopalan}},\
  }\href@noop {} {\  (\bibinfo {year} {2008})},\ \Eprint
  {http://arxiv.org/abs/0804.3858} {arXiv:0804.3858 [hep-ph]} \BibitemShut
  {NoStop}%
%%CITATION = 0804.3858;%%
\bibitem [{\citenamefont {Dumitru}\ \emph {et~al.}(2014)\citenamefont
  {Dumitru}, \citenamefont {McLerran},\ and\ \citenamefont
  {Skokov}}]{Dumitru:2014yza}%
  \BibitemOpen
  \bibfield  {author} {\bibinfo {author} {\bibfnamefont {A.}~\bibnamefont
  {Dumitru}}, \bibinfo {author} {\bibfnamefont {L.}~\bibnamefont {McLerran}}, \
  and\ \bibinfo {author} {\bibfnamefont {V.}~\bibnamefont {Skokov}},\
  }\href@noop {} {\  (\bibinfo {year} {2014})},\ \Eprint
  {http://arxiv.org/abs/1410.4844} {arXiv:1410.4844 [hep-ph]} \BibitemShut
  {NoStop}%
%%CITATION = ARXIV:1410.4844;%%
\bibitem [{\citenamefont {Adare}\ \emph {et~al.}(2013)\citenamefont {Adare}
  \emph {et~al.}}]{Adare:2013piz}%
  \BibitemOpen
  \bibfield  {author} {\bibinfo {author} {\bibfnamefont {A.}~\bibnamefont
  {Adare}} \emph {et~al.} (\bibinfo {collaboration} {PHENIX Collaboration}),\
  }\href {\doibase 10.1103/PhysRevLett.111.212301} {\bibfield  {journal}
  {\bibinfo  {journal} {Phys.Rev.Lett.}\ }\textbf {\bibinfo {volume} {111}},\
  \bibinfo {pages} {212301} (\bibinfo {year} {2013})},\ \Eprint
  {http://arxiv.org/abs/1303.1794} {arXiv:1303.1794 [nucl-ex]} \BibitemShut
  {NoStop}%
%%CITATION = ARXIV:1303.1794;%%
\bibitem [{\citenamefont {Wang}(2014)}]{Wang:2014qiw}%
  \BibitemOpen
  \bibfield  {author} {\bibinfo {author} {\bibfnamefont {F.}~\bibnamefont
  {Wang}} (\bibinfo {collaboration} {STAR Collaboration}),\ }\href {\doibase
  10.1016/j.nuclphysa.2014.09.063} {\  (\bibinfo {year} {2014}),\
  10.1016/j.nuclphysa.2014.09.063},\ \Eprint {http://arxiv.org/abs/1404.2674}
  {arXiv:1404.2674 [nucl-ex]} \BibitemShut {NoStop}%
%%CITATION = ARXIV:1404.2674;%%
\bibitem [{\citenamefont {Yi}(2014)}]{LiYiforSTAR:2014mea}%
  \BibitemOpen
  \bibfield  {author} {\bibinfo {author} {\bibfnamefont {L.}~\bibnamefont {Yi}}
  (\bibinfo {collaboration} {STAR Collaboration}),\ }\href {\doibase
  10.1016/j.nuclphysa.2014.10.008} {\bibfield  {journal} {\bibinfo  {journal}
  {Nucl.Phys.}\ }\textbf {\bibinfo {volume} {A}} (\bibinfo {year} {2014}),\
  10.1016/j.nuclphysa.2014.10.008},\ \Eprint {http://arxiv.org/abs/1410.1978}
  {arXiv:1410.1978 [nucl-ex]} \BibitemShut {NoStop}%
%%CITATION = ARXIV:1410.1978;%%
\bibitem [{\citenamefont {Abdelwahab}\ \emph {et~al.}(2014)\citenamefont
  {Abdelwahab} \emph {et~al.}}]{Abdelwahab:2014cvd}%
  \BibitemOpen
  \bibfield  {author} {\bibinfo {author} {\bibfnamefont {N.}~\bibnamefont
  {Abdelwahab}} \emph {et~al.} (\bibinfo {collaboration} {STAR
  Collaboration}),\ }\href@noop {} {\  (\bibinfo {year} {2014})},\ \Eprint
  {http://arxiv.org/abs/1410.3524} {arXiv:1410.3524 [nucl-ex]} \BibitemShut
  {NoStop}%
%%CITATION = ARXIV:1410.3524;%%
\bibitem [{\citenamefont {Huang}()}]{he3:2014is}%
  \BibitemOpen
  \bibfield  {author} {\bibinfo {author} {\bibfnamefont {S.}~\bibnamefont
  {Huang}} (\bibinfo {collaboration} {PHENIX Collaboration}),\ }\href@noop {}
  {\bibinfo  {journal} {Talk at IS2014, https://indico.cern.ch/event/336283/ \\
  session/16/contribution/6/material/slides/1.pdf}\ }\BibitemShut {NoStop}%
\end{thebibliography}%

\end{document}